\documentclass[12pt,twoside]{article}

\usepackage{a4}
\usepackage{epsf} 
\usepackage{cite}
\usepackage{colordvi}

\newcommand{ \be  }{\begin{equation}}
\newcommand{ \ee  }{\end{equation}} 
\newcommand{ \bea }{\begin{eqnarray}}
\newcommand{ \eea }{\end{eqnarray}}
\newcommand{ \beas }{\begin{eqnarray*}}
\newcommand{ \eeas }{\end{eqnarray*}}

\newcommand{ \NPB    }[3]{Nucl. Phys.      {\bf B#1} (#2) #3}
\newcommand{ \PLB    }[3]{Phys. Lett. B    {\bf #1}  (#2) #3}
\newcommand{ \PLBold }[3]{Phys. Lett.      {\bf #1B} (#2) #3}
\newcommand{ \PRD    }[3]{Phys. Rev. D     {\bf #1}  (#2) #3}

\newcommand{ \PREP   }[3]{Phys. Rep.       {\bf #1}  (#2) #3}

\newcommand{ \EPJC   }[3]{Eur. Phys. J. C  {\bf #1}  (#2) #3}

\newcommand{ \centeron }[2]{{\setbox0=\hbox{#1}\setbox1=\hbox{#2}\ifdim
                             \wd1>\wd0\kern.5\wd1\kern-.5\wd0\fi \copy0
                             \kern-.5\wd0\kern-.5\wd1\copy1\ifdim\wd0>\wd1
                             \kern.5\wd0\kern-.5\wd1\fi}}
\newcommand{ \ltap }{\;\centeron{\raise.35ex\hbox{$<$}}
                     {\lower.65ex\hbox{$\sim$}}\;}
\newcommand{ \gtap }{\;\centeron{\raise.35ex\hbox{$>$}}
                     {\lower.65ex\hbox{$\sim$}}\;}

\newcommand{ \lsim }{\mathrel{\ltap}}

\newcommand{ \slashchar }[1]{\setbox0=\hbox{$#1$}   
   \dimen0=\wd0                                     
   \setbox1=\hbox{/} \dimen1=\wd1                   
   \ifdim\dimen0>\dimen1                            
      \rlap{\hbox to \dimen0{\hfil/\hfil}}          
      #1                                            
   \else                                            
      \rlap{\hbox to \dimen1{\hfil$#1$\hfil}}       
      /                                             
   \fi}                                             %

\newcommand{ \NI         }{ \tilde{N}_1 }

\def\G{ \tilde{G} }

\newcommand{ \lR         }{ { \tilde \ell}_R }

\newcommand{ \tauu       }{ { \tilde \tau}_1 }

\newcommand{\sss}{\scriptscriptstyle}

\newcommand{\Z}{Z^{\sss 0}}

\newcommand{\epem}{e^{\sss +}e^{\sss -}}

\newcommand{\stopu}{\tilde{t}_1}
\newcommand{\stopd}{\tilde{t}_2}

\begin{document}
\pagestyle{myheadings} 
\markboth
{\it Ambrosanio, Mele, Petrarca, Polesello, Rimoldi / Measuring the SUSY Breaking...}
{\it Ambrosanio, Mele, Petrarca, Polesello, Rimoldi / Measuring the SUSY Breaking...}

\begin{titlepage}
\thispagestyle{empty} 
\null~\vspace{-2.7cm} 
\begin{flushright} 
hep-ph/0010081          \hfill  CERN-TH/2000-206
\end{flushright} 
\hrule\hfill

\vspace{-0.1cm}

\hrule\hfill

\vspace{0.2cm} 
                                        
\begin{center}
{\BrickRed{\Large \bf 
Measuring the SUSY Breaking Scale at the LHC}} \\
\vspace{0.2cm} 
{\BrickRed{\Large \bf 
in the Slepton NLSP Scenario of GMSB Models}}$^{\: \star}$

~\\

{\large {\bf S.~Ambrosanio}~$^{a}$,  
        {\bf B.~Mele}~$^{b,c}$, 
        {\bf S.~Petrarca}~$^{c,b}$,
        {\bf G.~Polesello}~$^{d}$,
        {\bf A.~Rimoldi}~$^{e,f}$ \\

~\\

$^a$ CERN, {\it Theory Division}, 
     CH--1211 Geneva 23, Switzerland  \\
~\\
$^b$ INFN, {\it Sezione di Roma I}, Roma, Italy \\
~\\
$^c$ {\it Dipartimento di Fisica, 
     Universit\`a ``La Sapienza''}, \\
     p.le Aldo Moro 2, I--00185 Roma, Italy \\
~\\
$^d$ INFN, {\it Sezione di Pavia}, via Bassi 6, I--27100 Pavia, Italy \\
~\\
$^e$ CERN, {\it EP Division}, 
     CH--1211 Geneva 23, Switzerland  \\
~\\
$^f$ {\it Dipartimento di Fisica Nucleare e Teorica, Universit\`a di Pavia}, \\
     via Bassi 6, I--27100 Pavia, Italy
}
\end{center}

\vspace*{\fill}  

\begin{center} 
{\bf Abstract} \\
\end{center}
{\small 
We report a study on the measurement of the SUSY breaking scale
$\sqrt{F}$ in the framework of gauge-mediated supersymmetry breaking 
(GMSB) models at the LHC. The work is focused on the GMSB scenario where
a stau is the next-to-lightest SUSY particle (NLSP) and decays into a
gravitino with lifetime $c\tau_{\rm NLSP}$ in the range 0.5~m to 1~km.
We study the identification of long-lived sleptons using the momentum and 
time of flight measurements in the muon chambers of the ATLAS experiment. 
A realistic evaluation of the statistical and systematic uncertainties 
on the measurement of the slepton mass and lifetime is performed, based on 
a detailed simulation of the detector response. 
Accessible range and precision on $\sqrt{F}$ achievable with a counting 
method are assessed. Many features of our analysis can be extended to
the study of different theoretical frameworks with similar signatures 
at the LHC. 
}

\vspace*{\fill}  

\noindent 
\parbox{0.4\textwidth}{\hrule\hfill} \\ 
{\small 
$^\star$\ To appear in
{\it The Journal of High Energy Physics}
} 

\end{titlepage}

\setcounter{page}{0} 

\thispagestyle{empty} 
~\\ 
\newpage 
\thispagestyle{plain} 

\section{Introduction}
\label{sec:intro}

\noindent 
Since no superpartner has been detected at collider experiments so far,
supersymmetry (SUSY) cannot be an exact symmetry of Nature.
The requirement of ``soft'' supersymmetry breaking (SSB)~\cite{StevePrimer} 
alone gives rise to a large number of  free parameters. 
Hence, motivated theoretical hypotheses on the nature of SSB and the mechanism
through which it is transmitted to the visible sector of the theory
-- here assumed to be the one predicted by the minimal SUSY extension of the
Standard Model (MSSM) --  are vital. 
If SUSY is broken at energies of the order of the Planck mass and the SSB 
sector communicates with the MSSM sector through gravitational interactions
only, one falls in the supergravity (SUGRA) scheme.
The most promising alternative to SUGRA is based instead on the hypothesis
that the SSB occurs at relatively low energy scales and it is mediated 
mainly by 
gauge interactions (GMSB)~\cite{oldGMSB,newGMSB,GR-GMSB}. 
This scheme
provides a natural, automatic suppression of the SUSY contributions 
to flavour-changing and CP-violating processes. 
Furthermore, in the simplest versions of GMSB the MSSM spectrum and 
other observables depend on just a handful of parameters, usually
chosen to be
\be
M_{\rm mess}, \; N_{\rm mess}, \; \Lambda, \; \tan\beta, \; {\rm sign}(\mu),
\label{eq:pars}
\ee
where $M_{\rm mess}$ is the overall messenger scale; $N_{\rm mess}$ is the 
so-called messenger index, parameterising the structure of the messenger
sector; $\Lambda$ is the universal soft SUSY breaking scale felt by the
low-energy sector; $\tan\beta$ is the ratio of the vacuum expectation 
values of the two Higgs doublets; sign($\mu$) (we use the convention of
Ref.~\cite{StevePrimer}) is the ambiguity left for the SUSY higgsino mass 
after imposing the conditions for a correct electroweak symmetry breaking 
(EWSB) (see e.g., Refs.~\cite{GMSBmodels1,GMSBmodels2,AKM-LEP2,AB-LC}).

The phenomenology of GMSB (and more in general of any theory with low-energy
SSB) is characterised by the presence of a very light gravitino 
$\G$ with mass~\cite{Fayet} 
\be 
m_{3/2} = m_{\G} = \frac{F}{\sqrt{3}M'_P} \simeq 
\left(\frac{\sqrt{F}}{100 \; {\rm TeV}}\right)^2 2.37 \; {\rm eV},  
\label{eq:Gmass}
\ee
where $\sqrt{F}$ is the fundamental scale of SSB and 
$M'_P = 2.44 \times 10^{18}$~GeV is the reduced Planck mass.
Since $\sqrt{F}$ is typically of order 100 TeV, the $\G$ is always the 
lightest SUSY particle (LSP) in these theories. Hence, if $R$-parity is 
conserved, any MSSM particle will decay into the gravitino. 
Depending on $\sqrt{F}$, the interactions 
of the gravitino, although much weaker than gauge and Yukawa interactions,
can still be strong enough to be of relevance for collider physics. 
In most cases the last step of any SUSY decay chain is 
the decay of the next-to-lightest SUSY particle (NLSP), which can  
occur either outside or inside a typical detector, possibly 
close to the interaction point. 
For particular ranges of lifetimes and assumptions on the NLSP
nature, the signature can be spectacular.

The typical NLSP lifetime for decaying into $\G$ is
\be 
\frac{c \tau_{\rm NLSP}}{\rm cm} \simeq \frac{1}{100 {\cal B}}
\left(\frac{\sqrt{F}}{100 \; {\rm TeV}}\right)^4 
\left(\frac{m_{\rm NLSP}} {100 \; {\rm GeV}}\right)^{-5},
\label{eq:NLSPtau}
\ee 
where ${\cal B}$ is a number of order unity depending mainly on the nature
of the NLSP.

The nature of the NLSP -- or, better, of the sparticle(s) having a large 
branching ratio (BR) for decaying into the gravitino and the relevant 
Standard Model (SM) partner -- determines four main scenarios giving rise 
to qualitatively different phenomenology:

\begin{description} 

\item[Neutralino NLSP scenario:] Occurs whenever 
$m_{\NI} < (m_{\tauu} - m_{\tau})$. Here typically a decay  
$\NI \to \G\gamma$ is the final step of decay chains following
any SUSY production process. As a consequence, the main inclusive signature 
at colliders is prompt or displaced photon pairs + X + missing energy,
depending on the $\NI$ lifetime. $\NI \to \G \Z$ and other minor channels 
may also be relevant at TeV colliders.    

\item[Stau NLSP scenario:] Realised if 
$m_{\tauu} < {\rm Min}[m_{\NI}, m_{\lR}] - m_{\tau}$, 
features $\tauu \to \G \tau$ decays, producing $\tau$ pairs (if $\tauu$ 
decays promptly) or charged semi-stable $\tauu$ tracks or decay kinks 
+ X + missing energy (for larger $\tauu$ lifetimes). 
Here $\ell$ stands for $e$ or $\mu$. 

\item[Slepton co-NLSP scenario:] When 
$m_{\lR} < {\rm Min}[m_{\NI}, m_{\tauu} + m_{\tau}]$, 
$\lR \to \G \ell$ decays are also open with large BR, since 
$\lR \to \ell\tauu^{\pm}\tau^{\mp}$ decays are kinematically forbidden. 
In addition to the signatures of the stau NLSP scenario, one also gets 
$\ell^+\ell^-$ pairs or $\lR$ tracks or decay kinks. 

\item[Neutralino-stau co-NLSP scenario:] If 
$| m_{\tauu} - m_{\NI} | < m_{\tau}$ and $m_{\NI} < m_{\lR}$, 
both signatures of the neutralino NLSP and stau NLSP scenario are present
at the same time, since $\NI \leftrightarrow \tauu$ decays are not allowed
by phase space. 

\end{description}

Note that in the GMSB parameters space $m_{\lR} > m_{\tauu}$ always. 
Also, one should keep in mind that the classification above is  
an indicative scheme valid in the limit $m_e$, $m_\mu \to 0$, neglecting 
also those cases where a fine-tuned choice of $\sqrt{F}$ and the sparticle 
masses may give rise to competition between phase-space suppressed decay 
channels from one ordinary sparticle to another and sparticle decays to the 
gravitino~\cite{AKM2}. 

The fundamental scale of SUSY breaking $\sqrt{F}$ is a crucial 
parameter for the phenomenology of a SUSY theory.
In the SUGRA framework, the gravitino mass sets the scale of the soft SUSY 
breaking masses ($\sim 0.1-1$ TeV), so that $\sqrt{F}$ is typically as large 
as $\sim 10^{10-11}$~GeV [cfr. Eq.~(\ref{eq:Gmass})].
As a consequence, the interactions of the $\G$ with the other MSSM
particles ($\sim F^{-1}$) are too weak to be of relevance
in collider physics and there is no direct way to access $\sqrt{F}$
experimentally. In GMSB theories the situation is different.
The soft SUSY breaking scale of the MSSM and the sparticle masses are set by
gauge interactions between the messenger and the low energy sectors
to be $\sim \alpha_{\rm SM}\Lambda$ [cfr. Eq.~(\ref{eq:bound}), next section],
 so that typical $\Lambda$ values are $\sim 10-100$ TeV. On the other hand,
$\sqrt{F}$ is only subject to a lower bound [cfr. Eq.~(\ref{eq:sqrtFmin}),
next section],
for which  values well below $10^{10}$~GeV and even as low as
several tens of TeV are reasonable.
$\G$ is in this case the LSP, and its interactions are strong enough to
allow NLSP decays into  $\G$ inside the typical detector size.
The latter circumstance gives us a chance for extracting $\sqrt{F}$
experimentally through a measurement of the NLSP mass and lifetime,
according to Eq.~(\ref{eq:NLSPtau}).

Furthermore, the possibility of determining $\sqrt{F}$ with good
precision opens a window on the physics of the SUSY breaking sector
(the so-called
``secluded'' sector) and the way this SUSY breaking is transmitted to the
messenger sector. Indeed, the characteristic scale of SUSY breaking felt
by the messengers (and hence by the MSSM sector) given by
$\sqrt{F_{\rm mess}}$ in Eq.~(\ref{eq:sqrtFmin}), next section, 
can be also determined
once the MSSM spectrum is known. By comparing the measured values
of $\sqrt{F}$ and $\sqrt{F_{\rm mess}}$ it might well be possible to
get information on the way the secluded and messenger sector communicate
to each other. For instance, if it turns out that $\sqrt{F_{\rm mess}}
\ll \sqrt{F}$, then it is very likely that the communication occurs
radiatively and the ratio $\sqrt{F_{\rm mess}/F}$ is given by some loop
factor. On the contrary, if the communication occurs via a direct
interaction, this ratio is just given by a Yukawa-type coupling
constant, with values $\ltap 1$, see Refs.~\cite{GR-GMSB,AKM-LEP2}.

An experimental method to determine $\sqrt{F}$ at a TeV scale $\epem$
collider through the measurement of the NLSP mass and lifetime was
presented in Ref.~\cite{AB-LC}, in the neutralino NLSP scenario.
Here, we are concerned with a similar problem at a hadron collider,
the LHC, and in the stau NLSP or slepton co-NLSP scenarios.
These scenarios are very promising at the LHC, providing signatures 
of semi-stable charged tracks coming from massive sleptons, therefore 
with $\beta$ significantly smaller than 1. In particular, we perform
our simulations in the ATLAS muon detector, whose large size and
excellent time resolution~\cite{TDR} allow a precision measurement of
the slepton time of flight from the production vertex out to the muon
chambers, and hence of the slepton velocity.
Moreover, in the stau NLSP or slepton co-NLSP scenarios, the
knowledge of the NLSP mass and lifetime is sufficient to
determine $\sqrt{F}$, since the factor ${\cal B}$ in
Eq.~(\ref{eq:NLSPtau}) is exactly equal to 1. This is not the
case in the neutralino NLSP scenario, where ${\cal B}$ depends
at least on the neutralino physical composition, and more information
and measurements are needed to extract a precise value of $\sqrt{F}$.

For this purpose, we generated about 30000 GMSB models under
well defined hypotheses, using a home-made program called 
{\tt SUSYFIRE}~\cite{SUSYFIRE}, as described in the following section. 

\section{GMSB Models}
\label{sec:models}

\noindent 
In the GMSB framework, the pattern of the MSSM spectrum is simple,
as all sparticle masses originate in the same way and scale 
approximately with a single parameter $\Lambda$, which sets
the amount of soft SUSY breaking felt by the visible sector. 
As a consequence, scalar and gaugino masses are related to each
other at a high energy scale, which is not the case in other SUSY 
frameworks, e.g. SUGRA. Also, it is possible to impose other 
conditions at a lower scale to achieve correct EWSB, and  
further reduce the dimension of the parameter space. 

To build our GMSB models, we adopted the usual phenomenological approach,
 following Ref.~\cite{AB-LC}.
We do not specify the origin of $\mu$, nor do we assume $B\mu = 0$ at 
the messenger scale. Instead, we impose correct EWSB to trade $\mu$
and $B\mu$ for $M_Z$ and $\tan\beta$, leaving the sign of $\mu$ undetermined. 
However, we recall that, to build a satisfactory GMSB model, one should also 
solve the latter problem in a more fundamental way, perhaps providing a 
dynamical mechanism to generate $\mu$ and $B\mu$, reasonably with values
of the same order of magnitude. This might be accomplished radiatively 
through some new interaction. However, in this case, the other
soft terms in the Higgs potential, namely $m^2_{H_{1,2}}$, will be also  
affected and this will in turn change the values of $|\mu|$ and $B\mu$ 
coming from EWSB conditions. 

To determine the MSSM spectrum and low-energy parameters, we solve 
the renormalisation group equation evolution with boundary conditions
at the $M_{\rm mess}$ scale, where
\bea
M_a & = & N_{\rm mess} \Lambda 
g\left(\frac{\Lambda}{M_{\rm mess}}\right) \alpha_a, 
\; \; \; (a=1, 2, 3) \nonumber \\
\tilde{m}^2 & =  & 2 N_{\rm mess} \Lambda^2 
f\left(\frac{\Lambda}{M_{\rm mess}}\right) 
\sum_a \left(\frac{\alpha_a}{4\pi}\right)^2 C_a,
\label{eq:bound}
\eea 
respectively for the gaugino and the scalar masses. 
The exact expressions for $g$ and $f$ at the one and two-loop level
can be found, e.g., in Ref.~\cite{AKM-LEP2},
and $C_a$ are the quadratic Casimir invariants for the scalar fields.
As usual, the scalar trilinear couplings $A_f$ are assumed to vanish
at the messenger scale, as suggested by the fact that they (and not
their square) are generated via gauge interactions with the messenger 
fields at the two loop-level only. 

 To single out the interesting region of the GMSB parameter space, 
we proceed as follows.
Barring the case where a neutralino is the NLSP and decays outside
the detector (large $\sqrt{F}$), the GMSB signatures are very spectacular
and the SM background is generally negligible or easily subtractable. 
With this in mind and being interested in GMSB phenomenology at the LHC, 
we consider only models where the NLSP mass is larger than 100~GeV, 
assuming that searches at LEP and Tevatron, if unsuccessful, will at the 
end exclude a softer spectrum in most cases. 
We require that $M_{\rm mess} > 1.01 \Lambda$, to prevent an excess
of fine-tuning of the messenger masses, and that the mass of the 
lightest messenger scalar be at least 10 TeV. We also impose 
$M_{\rm mess} > M_{\rm GUT} \; {\rm exp}(-125/N_{\rm mess})$,
to ensure the perturbativity of gauge interactions up to the 
GUT scale. Further, we do not consider models with $M_{\rm mess} \gtap 10^{5}
\Lambda$. As a result of this and other constraints, the messenger index 
$N_{\rm mess}$, which we assume to be an integer independent of the gauge 
group, cannot be larger than 8. To prevent the top Yukawa coupling from 
blowing up below the GUT scale, we require $\tan\beta > 1.2$
(this also takes partly into account the bounds from SUSY Higgs searches 
at LEP2).
Models with $\tan\beta \gtap 55$ (with a mild dependence on 
$\Lambda$) are forbidden by the EWSB requirement and typically fail 
to give $m^2_A > 0$.

The NLSP lifetime  is controlled  
by the fundamental SSB scale $\sqrt{F}$ value on a
model-by-model basis. Using perturbativity arguments, for each given 
set of GMSB parameters it is possible to determine a lower bound
according to~\cite{AKM-LEP2}
\be 
\sqrt{F} > \sqrt{F_{\rm mess}} = \sqrt{\Lambda M_{\rm mess}} > \Lambda.
\label{eq:sqrtFmin}
\ee
On the contrary, no solid arguments can be used to set an upper limit 
of relevance for collider physics, although some semi-qualitative
cosmological arguments are sometimes evoked.

To generate our model samples using {\tt SUSYFIRE}, we used logarithmic
steps for $\Lambda$ (between about 45 TeV/$N_{\rm mess}$ and about 220 
TeV/$\sqrt{N_{\rm mess}}$, which corresponds to excluding models with
sparticle masses above $\sim 4$ TeV), $M_{\rm mess}/\Lambda$ (between about 
1.01 and $10^5$) and $\tan\beta$ (between 1.2 and about 60), subject
to the constraints described above. 
{\tt SUSYFIRE} starts from the values of particle masses and gauge
couplings at the weak scale and then evolves them up to the messenger
scale through RGE's. At the messenger scale, it imposes the boundary 
conditions (\ref{eq:bound}) for the soft sparticle masses and then 
evolves the RGE's back to the weak scale. The decoupling of each 
sparticle at the proper threshold is taken into account. 
Two-loop RGE's are used for gauge couplings, third generation Yukawa
couplings and gaugino soft masses. The other RGE's are taken at the
one-loop level. At the scale $\sqrt{m_{\stopu}m_{\stopd}}$, EWSB conditions 
are imposed by means of the one-loop effective potential approach, 
including corrections from stops, sbottoms and staus. 
The program then evolves up again to $M_{\rm mess}$ and so on.
Three or four iterations are usually enough to get a good approximation
for the MSSM spectrum. 

\section{Setting the Example Points}
\label{sec:points}

\noindent
The two main parameters affecting the experimental measurement at the
LHC of the slepton NLSP properties are the slepton mass and momentum
distribution. Indeed, at a hadron collider most of the NLSP's come
from squark and gluino production, followed by cascade decays.
Thus, the momentum distribution is in general a function of the whole
MSSM spectrum. However, one can approximately assume that most of the
information on the NLSP momentum distribution is provided by the squark
mass scale $m_{\tilde q}$ only (in the stau NLSP scenario or slepton
co-NLSP scenarios of GMSB, one generally finds
$m_{\tilde{g}} \gtap m_{\tilde q}$).
To perform detailed simulations, we select a representative set of
GMSB models generated by {\tt SUSYFIRE}. We limit ourselves to models
with $m_{\rm NLSP} > 100$~GeV, motivated by the discussion in
Sec.~\ref{sec:models}, and $m_{\tilde q} < 2$ TeV, in order to yield
an adequate event statistics after a three-year low-luminosity run
(corresponding to 30 fb$^{-1}$) at the LHC.
Within these ranges, we choose eight points (four in the stau NLSP 
scenario and four in the slepton co-NLSP scenario), most of which 
representing extreme cases allowed by GMSB in the 
($m_{\rm NLSP}$, $m_{\tilde q}$) plane in order to cover the various 
possibilities.

\begin{table}
\begin{center}
\begin{tabular}{|r||r|r|r|r|c|} \hline
ID & $M_{\rm mess}$ (TeV) & $N_{\rm mess}$
   & $\Lambda$ (TeV)  & $\tan\beta$ & sign($\mu$)  \\ \hline\hline
1 & 1.79$\times 10^4$ & 3 &  26.6  &  7.22  & --   \\ \hline
2 & 5.28$\times 10^4$ & 3 &  26.0  &  2.28  & --   \\ \hline
3 & 4.36$\times 10^2$ & 5 &  41.9  & 53.7~  & +    \\ \hline
4 & 1.51$\times 10^2$ & 4 &  28.3  &  1.27  & --   \\ \hline
5 & 3.88$\times 10^4$ & 6 &  58.6  & 41.9~  & +    \\ \hline
6 & 2.31$\times 10^5$ & 3 &  65.2  &  1.83  & --   \\ \hline
7 & 7.57$\times 10^5$ & 3 & 104~~  &  8.54  & --   \\ \hline
8 & 4.79$\times 10^2$ & 5 &  71.9  &  3.27  & --   \\ \hline
\end{tabular}
\caption{\sl Input parameters of the example GMSB models chosen for our
study.}
\label{tab:tbg}
\end{center}
\end{table}
 
In Tab.~\ref{tab:tbg}, we list the input GMSB parameters we used to generate
these eight points, while in Tab.~\ref{tab:tbg1} we report the corresponding
values of the stau mass, the average squark mass $m_{\tilde q}$ 
and the gluino mass.
The ``NLSP'' column indicates whether the model belongs to the stau NLSP
or slepton co-NLSP scenario. The last column gives the total cross section
in pb for producing any pairs of SUSY particles at the LHC.

\begin{table}
\begin{center}
\begin{tabular}{|r||r|c|r|r|c|} \hline
ID &  $m_{\tilde\tau_1}$ (GeV)
   & ``NLSP'' & $m_{\tilde q}$ (GeV) & $m_{\tilde g}$ (GeV)
   & $\sigma$ (pb) \\ \hline \hline
1 &  100.1 & $\tauu$       &  577 &  631 & 42~~~~~ \\ \hline
2 &  100.4 & $\tilde\ell$  &  563 &  617 & 50~~~~~ \\ \hline
3 &  101.0 & $\tauu$       & 1190 & 1480 & ~0.59~  \\ \hline
4 &  103.4 & $\tilde\ell$  &  721 &  859 & 10~~~~~ \\ \hline
5 &  251.2 & $\tauu$       & 1910 & 2370 & ~0.023  \\ \hline
6 &  245.3 & $\tilde\ell$  & 1290 & 1410 & ~0.36~  \\ \hline
7 &  399.2 & $\tauu$       & 2000 & 2170 & ~0.017  \\ \hline
8 &  302.9 & $\tilde\ell$  & 1960 & 2430 & ~0.022  \\ \hline
\end{tabular}
\caption{\sl Features of the example GMSB model points studied 
($\tilde{\ell} = \tilde{e}_R$, $\tilde{\mu}_R$, $\tauu$).}
\label{tab:tbg1}
\end{center}
\end{table}

\begin{figure}
\centerline{
\epsfxsize = 0.6\textwidth
\epsfbox{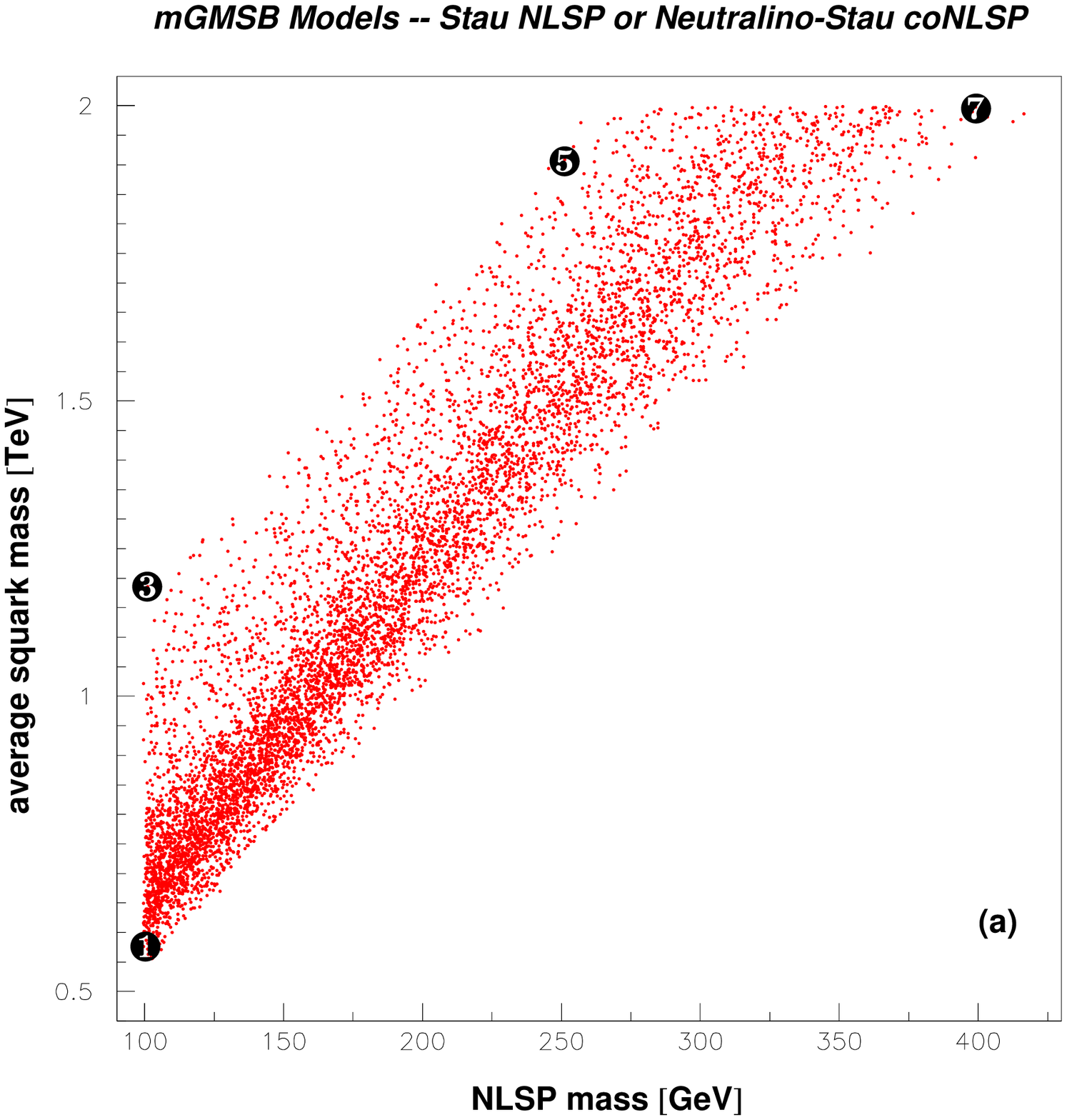}
\hspace{-0.5cm}
\epsfxsize = 0.6\textwidth
\epsfbox{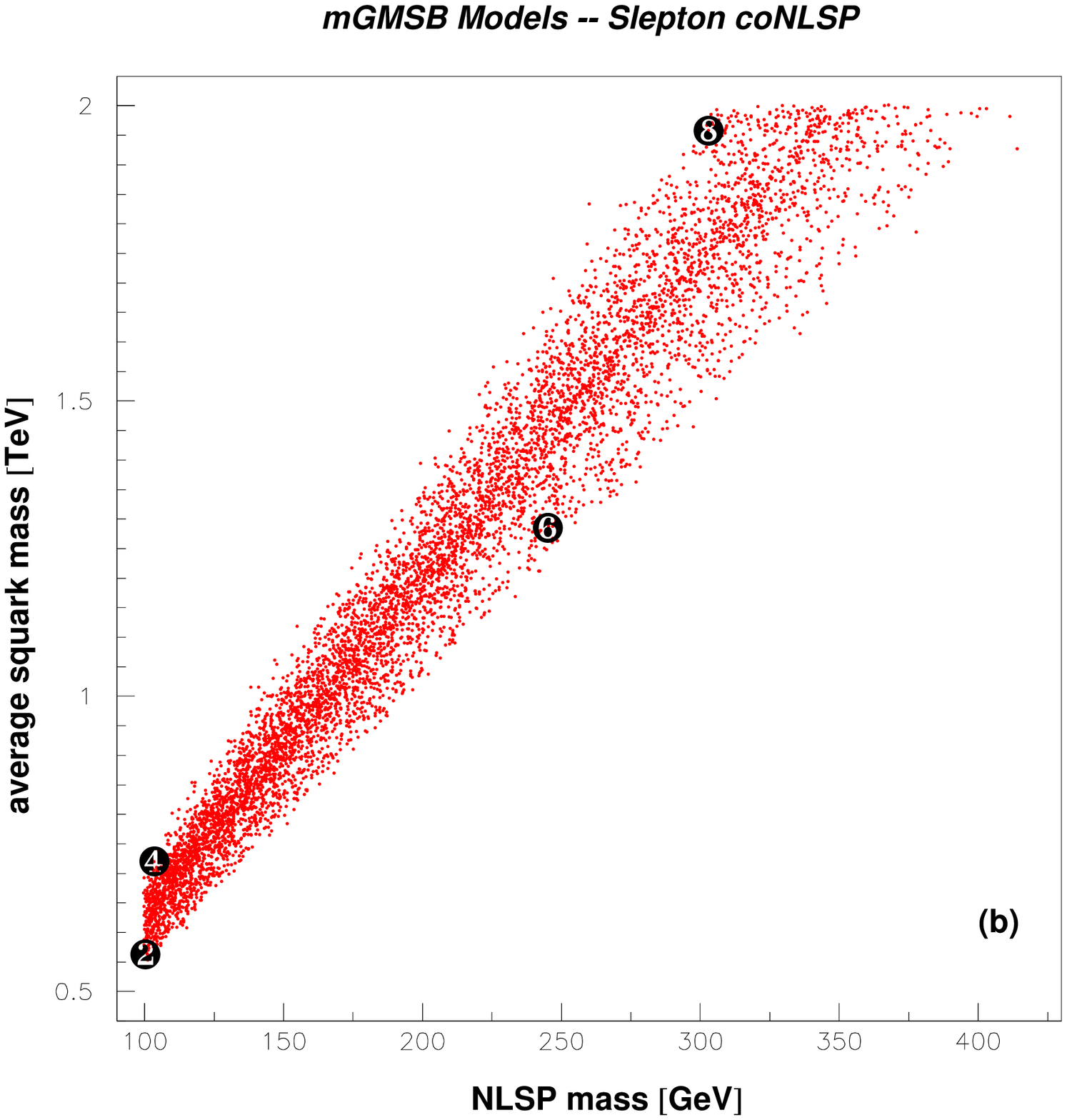}
}
\caption{\sl
Scatter plots in the ($m_{\rm NLSP}$, $m_{\tilde q}$) plane for all
relevant GMSB models generated. a) stau NLSP or neutralino-stau NLSP
scenarios; b) slepton co-NLSP scenario. The eight sample models 
of Tabs.~\ref{tab:tbg} and \ref{tab:tbg1} are highlighted with 
their reference number.}
\label{fig:models}
\end{figure}

The scatter plots in Fig.~\ref{fig:models} show our eight example points, 
together with all the relevant GMSB models we generated, in the 
($m_{\rm NLSP}$, $m_{\tilde q}$) plane.
In particular, all models where the charged tracks come from semi-stable 
$\tauu$'s only (i.e., stau NLSP or neutralino-stau co-NLSP scenarios) 
are displayed in Fig.~\ref{fig:models}a, while models in the
slepton co-NLSP scenario are shown in Fig.~\ref{fig:models}b.

For each sample model point, the events were generated with the 
{\tt ISAJET} Monte Carlo \cite{isajet} that incorporates the calculation 
of the SUSY mass spectrum and branching fraction
using the GMSB parameters as input. We have checked that for
the eight model points considered the sparticle masses
calculated with {\tt ISAJET} are in good agreement with the output
of {\tt SUSYFIRE}.
The generated events were then passed through {\tt ATLFAST} \cite{ATLFAST},
a fast particle-level simulation of the ATLAS detector. The {\tt ATLFAST}
package, was only used to evaluate the efficiency of the calorimetric
trigger that selects the GMSB events and of the event selection 
cuts. The detailed response of the detector
to the slepton NLSP has been parametrised for this work using the results
of a full simulation study, as described in the next section.

\section{Slepton Detection}
\label{sec:slepdet}

\noindent
The experimental  signatures of heavy long-lived charged
particles  have  been  discussed both in the framework
of GMSB and in more general scenarios \cite{leandro,drtata,femoroi}.
The two main observables which can be used to separate
these particles from muons are the high specific ionisation
and the  time of flight in the detector.

We concentrate here on the measurement of the time of flight
which is made possible by the timing precision ($\lsim 1$~ns)
and by the size of the ATLAS muon spectrometer.
In the barrel part of the detector ($|\eta|<1$) the precision
muon system consists of three multilayers of precision drift tubes
immersed in a toroidal air-core magnetic field.
The three measuring  stations are located at distances
of approximately 5, 7.5 and 10 meters
from the interaction point. A particle crossing a drift chamber
ionises the chamber gas along its path, and the
electrons produced by the ionisation drift to the anode wire
under the influence of an electric field. The particle position
is calculated from the  measurement of the drift time of the
ionisation electrons to the anode wire.
In order to perform this calculation
a starting time $t_0$ for counting the drift time is needed, corresponding
to the time of flight of the particle from the production
point to the measuring station.
For a particle travelling  approximately at the speed of light, as a muon,
the $t_0$'s for the measuring stations are parameters of the detector geometry
and of the response of the front-end electronics \cite{leandro1}.
For a heavy particle the $t_0$ is a free
parameter,  function of the $\beta$ (= v/c) of the particle.
It was demonstrated with a full simulation of the ATLAS muon detector \cite{gpar}
that the $\beta$ of a particle can be measured by adjusting the
$t_0$ for each station in such a way so as to
minimise the $\chi^2$ of the reconstructed muon track.
The resolution on $\beta$  obtained  in \cite{gpar}  can be  approximately
parametrised as:
$ \sigma(\beta)/\beta^2=0.028$, and the resolution on the transverse
momentum measurement is comparable to the one expected for
muons.
We have therefore simulated the detector response to NLSP sleptons
by smearing the slepton momentum and $\beta$  according
to the parametrisations in \cite{gpar}. The full simulation has only been
performed for particles produced centrally in the detector
($|\eta|\sim0$). We conservatively use the parametrisation
used in those conditions for the full $|\eta|$ coverage of the detector.
In fact, for all other  pseudorapidities the flight path
from the interaction point to the first station, and the distance
among measuring stations is larger than for $\eta=0$, and therefore
the resolution on $\beta$ is expected to improve.

\section{Triggering on GMSB Events}
\label{sec:trigger}

\noindent
In order to evaluate the available statistics for slepton mass
and lifetime measurements, we need to evaluate the trigger efficiency for 
the SUSY events.

The trigger system of the ATLAS experiment is described in detail
in \cite{TDR}.  Three levels of trigger are envisaged.  The first level
is exclusively hardware, and the information from the muon  detectors
and from the calorimeters are treated separately.  The second level refines
the first level by connecting the information from different detectors.
Finally the third level, also called 'event filter', applies the
full off line reconstruction algorithm to the data.

The quasi-stable NLSP events  can be selected in 
the ATLAS detector by  using either the muon or the calorimetric trigger. 
 
The first option has been studied in a preliminary way in Ref.~\cite{leandro}. 
An approximate evaluation for particles with pseudorapidity 
$|\eta|<1$ gives an efficiency of 50\% for $\beta=0.5$, increasing 
to basically full efficiency for $\beta=0.7$ for the trigger coincidence
based on high $P_T$ muons ($P_T>20$~GeV). 
No comparable study exists for particles with $|\eta|>1$. In this case, 
due to the larger distance of the trigger station, the $\beta$ 
threshold for the trigger to be sensitive to the heavy sleptons
will be higher than in the case of the  particles produced centrally
in the detector. In summary, while it is sure that part of the heavy sleptons
will be accepted by the ATLAS muon trigger, at the present level of 
studies it is difficult to quote an efficiency for such a trigger, 
especially in the region of low $\beta$, which 
is the one yielding the best resolution for the 
slepton mass measurement.

In minimal GMSB models with slepton masses larger than 100~GeV, the
squark masses are larger than 500~GeV.
Therefore the events with production of strongly interacting sparticles 
will in general contain multiple high $P_T$ jets.
As already observed in \cite{ihfp}, if the NLSP are visible, 
the $P_{T}^{\rm miss}$~is generated by the neutrinos
in the cascade decays, and its spectrum is relatively soft. 
If the $P_{T}^{\rm miss}$~is calculated only from the energy deposit in 
the calorimeter, neglecting the NLSP's and the muons,
the spectrum is much harder, and we recover the classical SUSY signature
of  $P_{T}^{\rm miss} +$jets.
The first level $P_{T}^{\rm miss}$~ trigger 
is based on the requirement of a jet with $P_T>50$~GeV and  
$P_{T}^{\rm miss} > 50$~GeV 
(both raised to 100~GeV for the high luminosity running),
where the $P_{T}^{\rm miss}$~is calculated exclusively from the energy 
deposit in all the calorimeter cells, and will therefore have a high 
efficiency for the models we are studying. 
Neglecting the detector smearing of the trigger thresholds,
the efficiency of the $P_{T}^{\rm miss}$~trigger for the 8 example points 
is of order 90\%.
A drawback of this purely calorimetric approach is the fact that processes
with low hadronic activity, such as direct slepton production  and direct
gaugino production are not selected. This appears clearly from the behaviour 
of the efficiency which is lower in the case of heavier squarks, when the 
fraction of events with direct gaugino production is higher. 
A part of these events will however be selected by the muon trigger. 

It has been estimated \cite{TDR} that the second level trigger will give an 
acceptable rate even if the  transverse momenta of triggering muons is not 
added to  the missing transverse momentum calculation.
Finally the full reconstruction at the event filter level will be able to 
select the SUSY events by applying to the muon detector the slepton 
reconstruction algorithm described above. The most likely scenario is that 
the SUSY events will be triggered 
by a combined use of the muon and the calorimetric triggers,
yielding an efficiency which is higher than the bare 
calorimeter efficiency. In the following, we will conservatively 
calculate the statistical errors on the measurements using the efficiencies 
of the calorimetric trigger.

\section{Event Selection}
\label{sec:cut}

\noindent
The SM backgrounds for the considered models are
processes with muons in the final state,
such as the production of $\bar tt$, $\bar bb$, $W$+jets, $Z$+jets, 
where a muon is misidentified as a slepton.\\
In order to extract a clean GMSB signal from the SM background
both tight identification criteria on 
a slepton candidate and kinematical cuts on the 
event  structure are needed.

\begin{figure}
\centerline{
\epsfxsize = 0.5\textwidth
\epsfbox{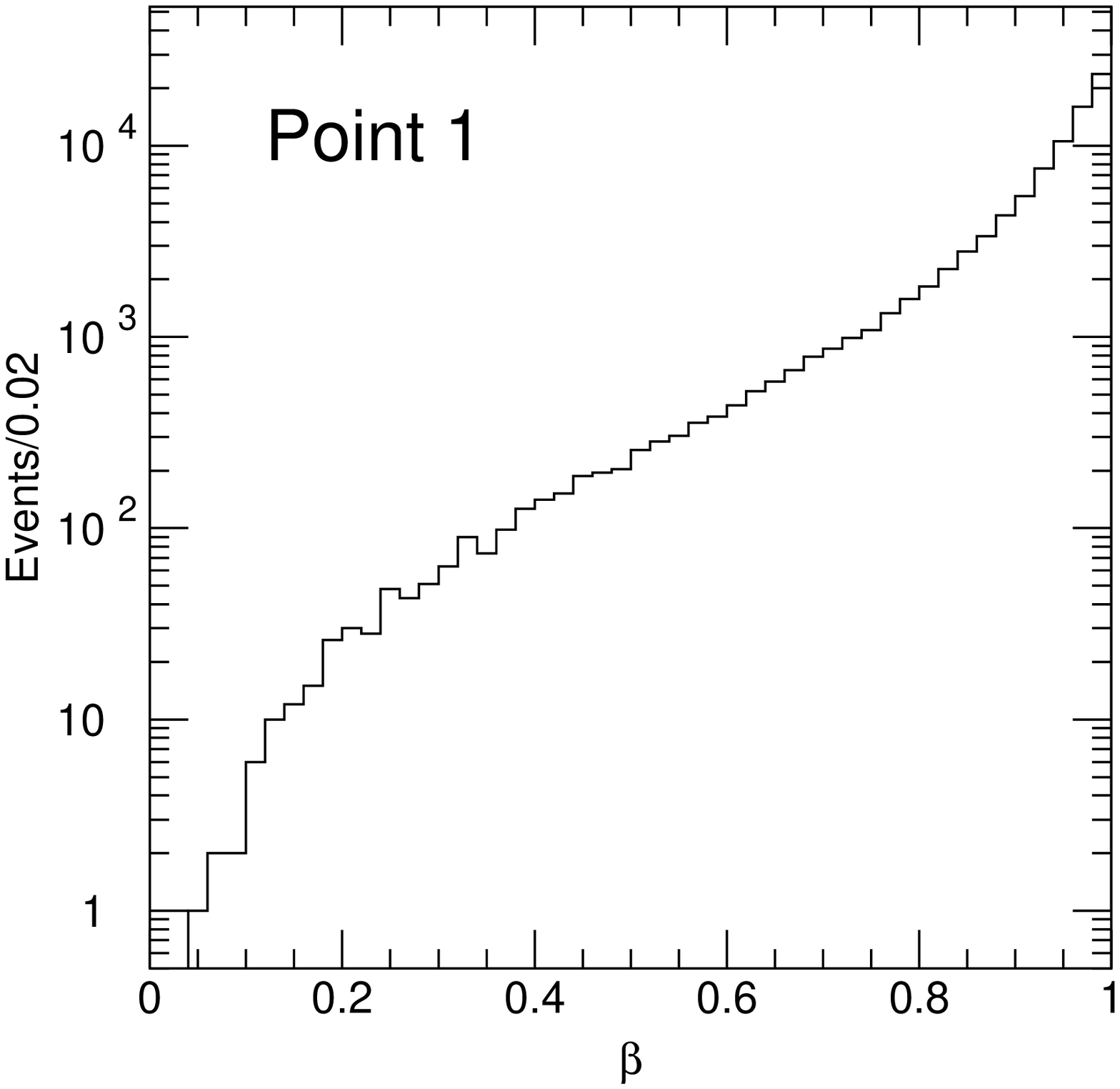}
\hfill
\epsfxsize = 0.5\textwidth
\epsfbox{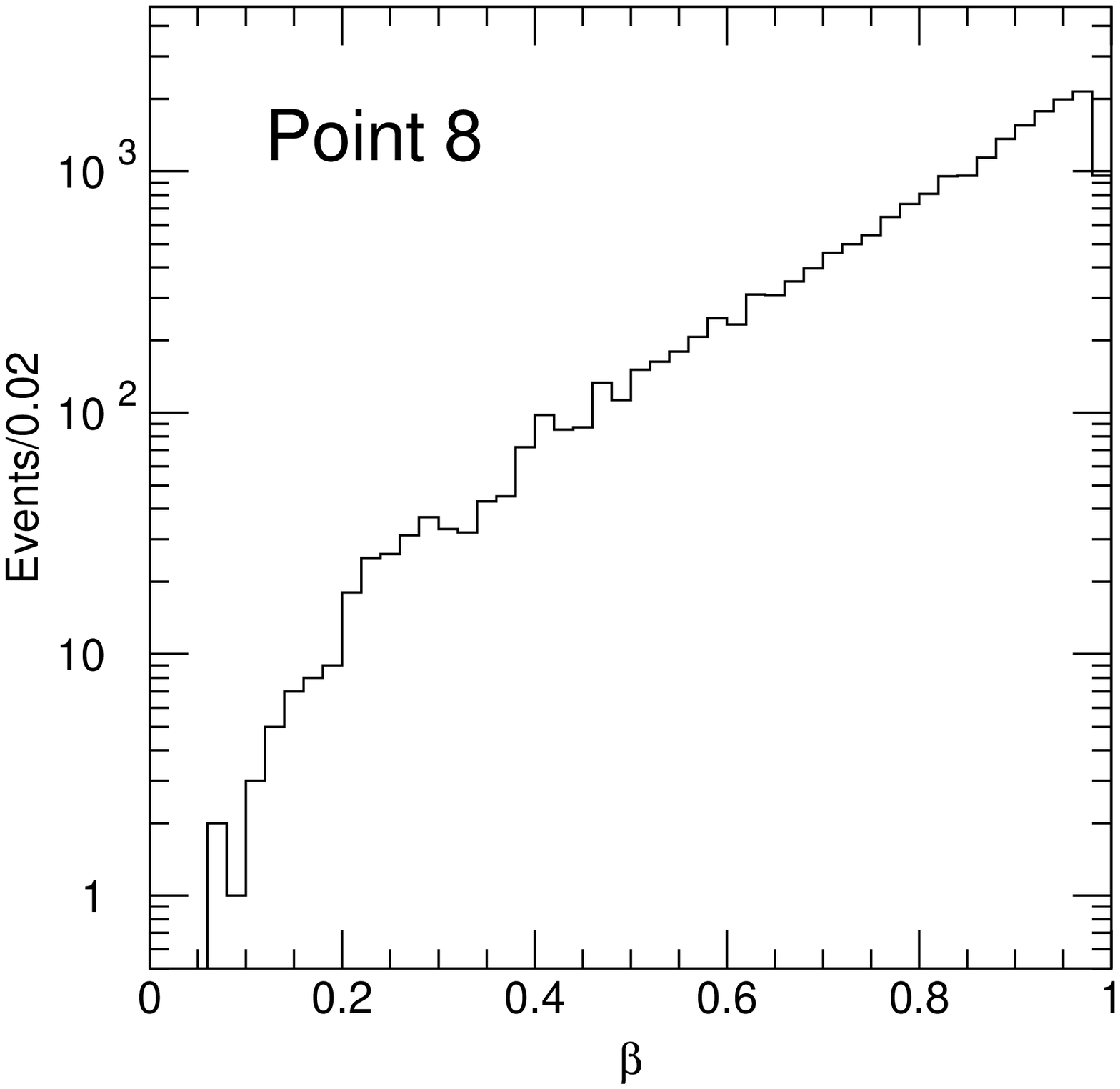}
}
\caption{\sl
Distribution of $\beta$ for the NLSP sleptons at Points 1 and 8.}
\label{fig:betadis}
\end{figure}

In order to select the heavy slepton, 
the key requirement is the presence of a track in the muon detector
with a measured $\beta$ ($\beta_{meas}$) different from one. 
Since the resolution on the 
$\beta$ measurement is $\sim$0.03, by requiring \mbox{$\beta_{meas}<0.91$} 
a rejection factor of $\sim$1000 is obtained on background muons
which have $\beta=1$. 
The $\beta$ distribution for sleptons is shown in
Fig.~\ref{fig:betadis} for model points 1 and 8. In all considered cases
a significant fraction of the events passes the  
\mbox{$\beta_{meas}<0.91$} cut.
Additional background rejection  is obtained by comparing 
the momentum and the $\beta$ of the track.
A track in the muon system is accepted as a slepton
candidate if it satisfies the following requirements:
\begin{itemize}
\item
$|\eta|<$2.4, to ensure that the particle is in the
acceptance of the muon trigger chamber, and therefore 
both transverse and longitudinal component of the momentum can be measured;
\item
$P_T>10$~GeV
after taking into account the energy loss in the calorimeters, 
to ensure that the particle traverse all the muon stations.
\item
It is isolated, where the isolation 
consists in requiring a total energy $<10$~GeV in the inner detector and 
in the calorimeter 
not associated with the candidate track in a pseudorapidity-azimuth  cone
$\Delta R \equiv \Delta\eta\times\Delta\phi=0.2$ 
around the track direction. 
\item
$\beta_{\rm meas}<0.91$, where $\beta_{\rm meas}$ is the $\beta$ of the
particle measured with the time-of-flight in the precision chambers;
\item
The momentum $P_{\rm meas}$ and  $\beta_{\rm meas}$ should be compatible with 
the slepton mass $m_{\tilde \ell}$, corresponding to the cuts:
$$
\frac{\beta_{\rm meas}-0.05}{\sqrt{1-(\beta_{\rm meas}-0.05)^2}} 
<  \frac{P_{\rm meas}}{m_{\tilde \ell}} <
\frac{0.91+0.05}{\sqrt{1-(0.91+0.05)^2}}  
$$
\end{itemize}

\begin{figure}
\centerline{
\epsfxsize = 0.5\textwidth
\epsfbox{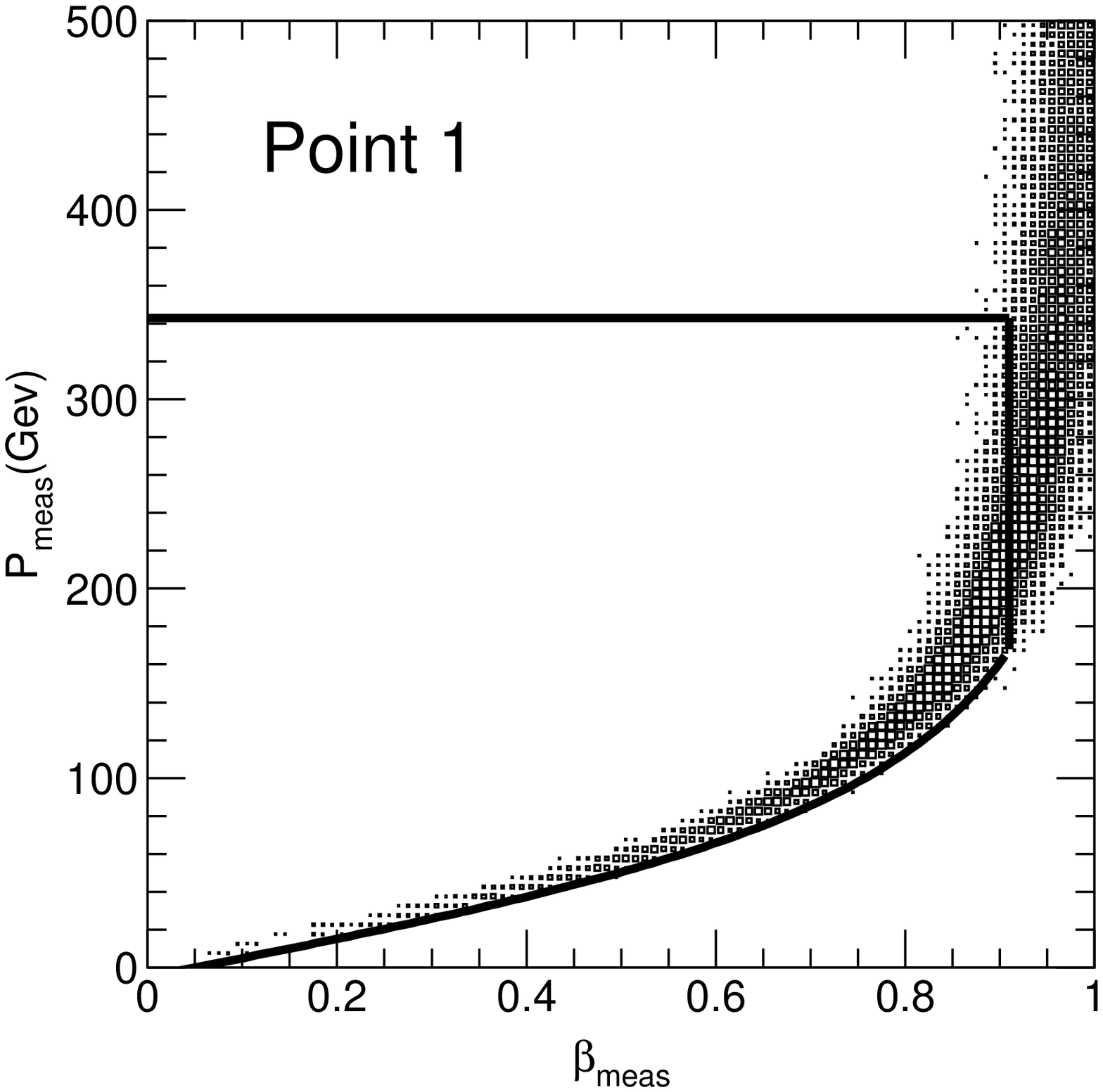}
\epsfxsize= 0.5\textwidth
\epsfbox{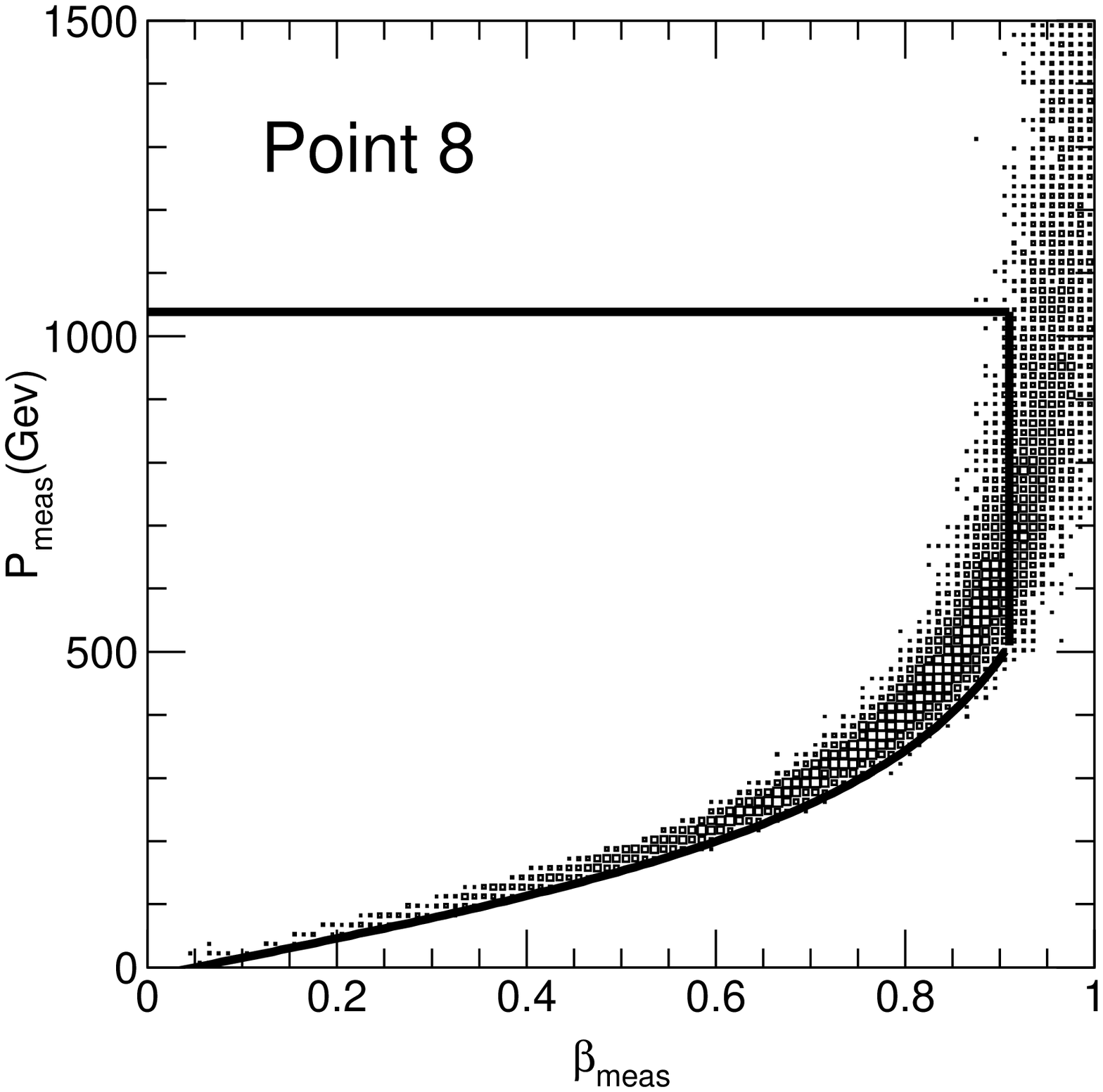}
}
\caption{\sl
Distribution of $P_{\rm meas}$ versus measured $\beta_{\rm meas}$ 
for sleptons, for Points 1 and 8. The region to the left 
of the thick lines in the ($\beta_{\rm meas}$, $P_{\rm meas}$) plane 
defines the selected slepton candidates}
\label{fig:betareg}
\end{figure}

The region in the ($\beta_{\rm meas}$, $P_{\rm meas}$) plane defined by the 
last two cuts is shown for Points~1 and 8 in Fig.~\ref{fig:betareg}.
With these cuts, the loss in slepton candidates compared to what one gets 
after the cut $\beta_{\rm meas}<0.91$ only is at the few percent level, 
with a significant gain in muon rejection, as only muons 
within a restricted momentum range can be misidentified as sleptons. 
For the lowest values of $m_{\tilde \ell}$ considered, the upper 
limit on $P_{\rm meas}$ is essential to reject the background 
from $W$+jets and $Z$+jets production. In fact, given the low 
jet multiplicity for this processes, the events passing the kinematic
selection described below typically contain muons with  
a few hundred GeV momentum.

The isolation requirement is necessary to reduce the background
from semileptonic decays of heavy quarks.
A possible additional rejection on muons using the ATLAS Transition Radiation 
Tracker is not considered in this analysis.

The trigger requirement and the presence of a slepton candidate 
already yield a significant GMSB signal over the SM background
from IVB + jets and top production. An overwhelming background 
is however expected from QCD production of $b$ jets, given the very soft 
kinematic requirements.
The squark mass scale for the GMSB models considered ranges between 
500~GeV and 2~TeV, giving a much larger transverse energy 
deposition in the calorimeter than  the QCD background.
To exploit this feature,  we build  an $m_{\rm eff}$ variable defined as:
$$
m_{\rm eff}=
\sum_{i=1}^{min(4,N_{jet})}P_T^{jet,i}+\sum_{i=1}^{min(2,N_{\mu})}P_T^{\mu,i},
$$
where $\mu$ is a track reconstructed in the muon detector, including the 
slepton candidates. 
This variable is similar to the one used for SUGRA 
inclusive studies in \cite{TDR}, but also takes into account the presence of 
final state sleptons, and has a high efficiency also for SUSY 
events with no squark/gluino production.

The final requirements for GMSB event selection are therefore:
\begin{itemize}
\item
at least one hadronic jet with $P_T>50$~GeV and a calorimetric 
\mbox{$E_T^{\rm miss}>50$~GeV} (trigger requirement);
\item
at least one slepton candidate as defined above; 
\item
$m_{\rm eff}>800$~GeV;
\end{itemize} 
For this indicative study, the cut 
on $m_{\rm eff}$ is set to a common value for all models, 
and the choice  is aimed at reducing the SM background to 
a few percent of the signal for the models with lowest statistics.

In order to study the SM background, approximately 1 million 
events for each of the following processes: $\bar tt$, $W$+jets, $Z$+jets, 
$WW$, $WZ$, and $\sim$2~million QCD events (in different
bins of $P_T$) were generated with the {\tt PYTHIA} Monte Carlo \cite{PYTHIA}. 
The number of expected events after cuts for the eight GMSB models and
for the main SM backgrounds for 
an integrated luminosity of 30~fb$^{-1}$ are given in Tab.~\ref{tab:rate}.

\begin{table}
\begin{center}
\begin{tabular}{|l||r|r|r|r|r|r|}
\hline
Model & Signal & $W$+Jets & $Z$+Jets & $\bar tt$ & QCD & Total BKGD \\
\hline
\hline
1 &    452163     &    9.6   &      6.8  &       5.3   &      8.0   &     29.7 \\
2 &    528420     &    9.6   &      6.9  &       5.3   &      8.0   &     29.9 \\
3 &      7437     &    9.5   &      6.9  &       5.3   &      8.0   &     29.9 \\
4 &    147354     &    9.5   &      7.1  &       5.6   &      7.4   &     29.6 \\
5 &       365     &    2.4   &     11.1  &       6.2   &      3.1   &     22.8 \\
6 &      6535     &    2.4   &     11.0  &       6.5   &      3.1   &     23.0 \\
7 &       326     &    1.0   &      5.9  &       3.4   &      0.5   &     10.8 \\
8 &       378     &    1.8   &      8.7  &       4.9   &      2.0   &     17.4 \\
\hline
\end{tabular}
\end{center}
\caption{\sl Number of events expected after cuts for the eight
example models and for the major background sources. 
The assumed integrated luminosity is 30~fb$^{-1}$. }
\label{tab:rate}
\end{table}

A number of signal events ranging from a few hundred for the models with
the 2~TeV squark mass scale to  a few hundred thousand for a 500~GeV
mass scale survive these cuts.  The corresponding 
background is of the order of a few tens of events, yielding
a very pure sample which can be used for measuring the
NLSP properties. The effect of a possible finite lifetime
on the event statistics will be addressed in detail when
studying the NLSP lifetime. 

\section{Slepton Mass Measurement}
\label{sec:mass}

\begin{figure}
\centerline{
\epsfxsize = 0.5\textwidth
\epsfbox{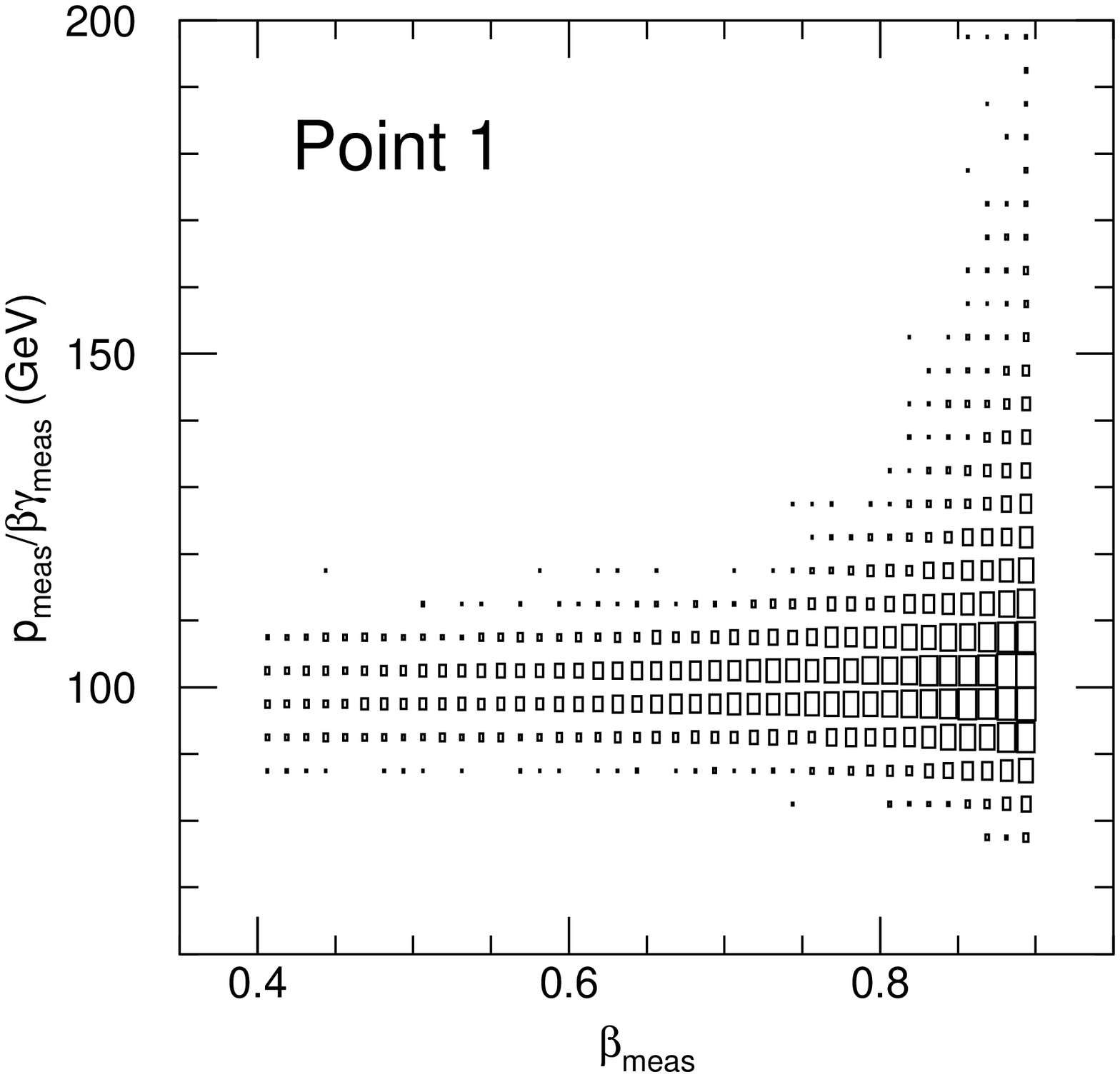}
\epsfxsize = 0.5\textwidth
\epsfbox{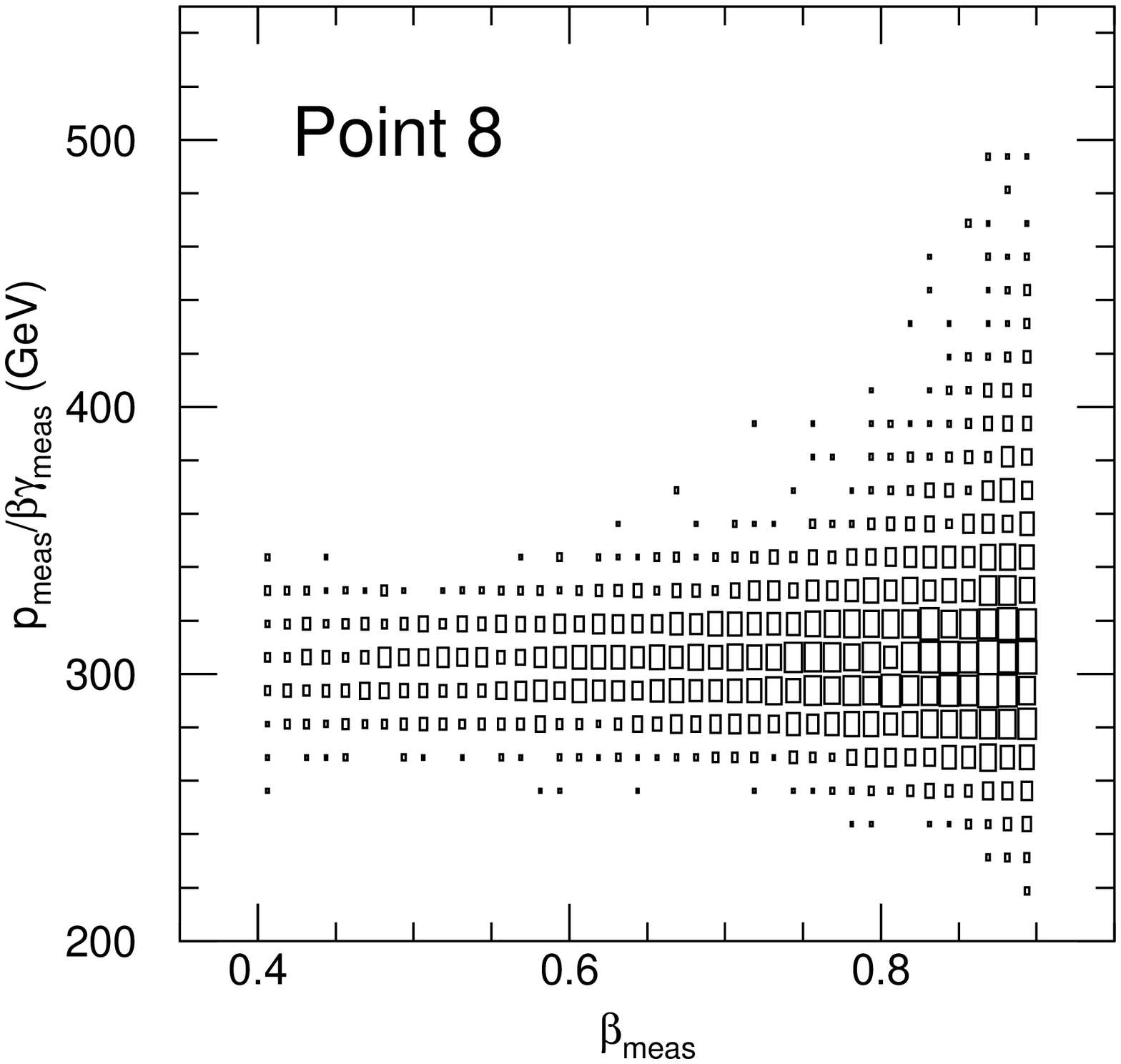}
}
\caption{\sl 
Distribution of the measured mass values for Points 1 and 8  as a function of
$\beta$. The number of events in the scatter plot is arbitrary.}
\label{fig:cor}
\end{figure}

\noindent
In order to perform this measurement, the particle momentum is needed.
The precision chambers only provide a measurement 
of the momentum components  transverse to the beam axis, so 
a measurement of the slepton pseudorapidity is needed. 
This can be performed either by a match with a track
in the inner detector, or using the information from the muon trigger chambers.
The first option requires a detailed study of the matching procedure between 
detectors.
This study was performed for muons in \cite{TDR}, but the results can not be
transferred automatically to the case of  heavy particles for which
the effect of multiple scattering in traversing the calorimetric system
is much more severe. 

In the case of the trigger chambers, as already discussed above, 
a limited time window around the beam crossing is read out, 
restricting the $\beta$ range for which the momentum can be measured.
We therefore evaluated the statistical precision achievable 
for the eight example models in two different $\beta$ intervals:
 $0.6<\beta<0.91$ and $0.8<\beta<0.91$.
For this measurement the sleptons are assumed to be stable. 

From the measurements of the slepton momentum and of the particle $\beta$,
the mass can be measured using the standard relation
$m=p\frac{\sqrt{1-\beta^2}}{\beta}$.
For each value of $\beta$ and momentum, the measurement error is
known, and it is given by the parametrisations in Ref.~\cite{gpar}.
Therefore, the most straightforward way to measure the mass
is just to use the weighted average of all the masses
calculated with the above formula. For two of the example points, we
show in Fig.~\ref{fig:cor} the distribution of the measured mass as 
a function of $\beta$. For values of $\beta$ in the range (0.6,0.9) the
spread in the mass measurement goes from $\sim5\%$ to
$\sim15\%$, rising with $\beta$ \cite{gpar}.
If enough statistics at low $\beta$ is available, the measurement
precision will be dominated by the events  with a $\beta$ value
below $\approx0.8$.
\begin{table}
\begin{center}
\begin{tabular}{|l||r|r|r|r|r|}
\hline
Model &  $m_{\rm NLSP}$ & $N_{\rm NLSP}$ & $\sigma_m$ (GeV) &  
$N_{\rm NLSP}$ & $\sigma_m$ (GeV)  \\
& (GeV)    & \multicolumn{2}{c|} {$0.6<\beta<0.91$} &  
\multicolumn{2}{c|}  {$0.8<\beta<0.91$}  \\ \hline\hline
 1 & 100.1 & 365047 & 0.010   & 246809 & 0.017 \\
 2 & 100.4 & 425790 & 0.0093  & 289020 & 0.016 \\
 3 & 101.0 & 5933   & 0.084   & 3940   & 0.14  \\
 4 & 103.4 & 125220 & 0.018   & 81876  & 0.031 \\
 5 & 251.2 & 335    & 0.92    & 214    & 1.6   \\
 6 & 245.3 & 5595   & 0.22    & 3675   & 0.37  \\
 7 & 399.2 & 312    & 1.7     & 192    & 3.0   \\
 8 & 302.9 & 408    & 1.0     & 249    & 1.9   \\
\hline
\end{tabular}
\end{center}
\caption{\sl Statistical errors on the NLSP mass measurement for the eight
example models. The assumed integrated luminosity is 30~fb$^{-1}$. }
\label{tab:num}
\end{table}

In Tab.~\ref{tab:num}, we show the numbers of the NLSP  candidates for an 
integrated luminosity of 30~fb$^{-1}$ and the expected errors on the mass 
measurement. Only the statistical errors are shown. The main systematic
error on this measurement will be the uncertainty 
on the NLSP momentum scale, as the systematic error on the
time measurement is already included in the parametrisation. 
We expect this uncertainty to be of order 0.1\% as for the muons,
if the accuracy of the energy scale measurement can be propagated
to the high momentum scale considered in this analysis.

From the numbers in the table, one expects that if the NLSP is long lived,
the measurement error on the NLSP mass will be dominated by the systematic 
error for models with a squark mass scale up to $\sim$1~TeV.

In conclusion, the slepton mass can be measured 
with a precision of a few permille for all the considered models 
with an integrated luminosity of 30~pb$^{-1}$.

\section{Slepton Lifetime Measurement}
\label{sec:lifetime}

\noindent
The measurement of the NLSP lifetime was studied in detail in 
\cite{AB-LC} for the case of a $\NI$ NLSP at a high energy $e^+e^-$ collider. 
In that work a number of methods were discussed  in the framework
of an idealised detector. By combining the different approaches
a wide range in NLSP lifetimes could be covered.\par
The aim of this study is to perform a realistic evaluation,
including the most important experimental effects, based on the
detailed simulation of the response of a real detector  
which is already in the construction phase.
For this reason we do not attempt to combine different methods,
each of which would require a dedicated detector study,
but we concentrate on a statistical method, which is based 
on the detailed study of the time of flight measurement 
capabilities of the ATLAS muon detector described in the previous sections.

We exploit the fact the two NLSP  are produced in each event. 
One  can therefore select events in which a slepton is detected 
through the time-of-flight measurement described above, 
and count in how many of these a second slepton candidate
is found. The ratio of the  number of events containing
two slepton candidates to the number of events with at least one 
candidate is a function of the slepton lifetime.
This measurement is in principle very simple, in practice it
requires an excellent control of the experimental 
sources of inefficiency for the detection of the second slepton.

The discussion in this section is therefore organised in a series
of logical steps.
We first describe the principle of the method, calculating 
the dependence of the measured ratio on the slepton lifetime, without
bothering how with real data it will be possible to connect the 
two quantities. 
From this study we estimate the achievable statistical
error for the considered models, and we evaluate how a systematic
uncertainty on the measured ratio propagates to the lifetime measurement. 
We then turn to analysing the main uncertainty sources,
including the presence of background from SUSY events, 
the incomplete knowledge of the underlying SUSY model
and the uncertainty on the detector acceptance. 
As a result of this analysis we estimate the range
of systematic uncertainties for the experimental measurement
of the lifetime. 
Based on this estimate in the next section we determine 
the achievable precision on the SUSY breaking scale $\sqrt{F}$ 
after the first three years of data-taking at the LHC.

\subsection{The Statistical Method}
\label{subsec:stat}

\noindent
We define $N_1$ the number of events passing 
the cuts discussed in Sec.~\ref{sec:cut}, 
with the additional requirement that there be 
at least one candidate slepton 
at a distance from the interaction vertex $>10$~m.
For the events thus selected 
we define  $N_2$ as the number of events where a second
track with a transverse momentum in excess of 10~GeV
is reconstructed in the muon system. The search for the
second particle should be as inclusive as possible, to minimise
the corrections which should be applied to the ratio. In particular
no slepton isolation is required, and
the tight cuts in the ($P_{\rm meas}, \beta_{\rm meas}$) plane
shown in Section \ref{sec:cut} are replaced by the requirement:

\begin{equation}
P_{\rm meas}>m_{\tilde \ell} \;
\frac{\beta_{\rm meas}-0.1}{\sqrt{1-(\beta_{\rm meas}-0.1)^2}},
\label{eq:2lep}
\end{equation}
where $m_{\tilde \ell}$ is the slepton mass.

The loss in signal for this cut is less than 0.1\%, thus introducing
a negligible uncertainty in the measurement, 
and the low momentum muons in the SUSY 
sample are rejected. The surviving background 
of high  momentum muons can be statistically
subtracted and it will be discussed in the following section.

\begin{figure}[htb]
\centerline{
\epsfxsize=0.7\textwidth
\epsffile{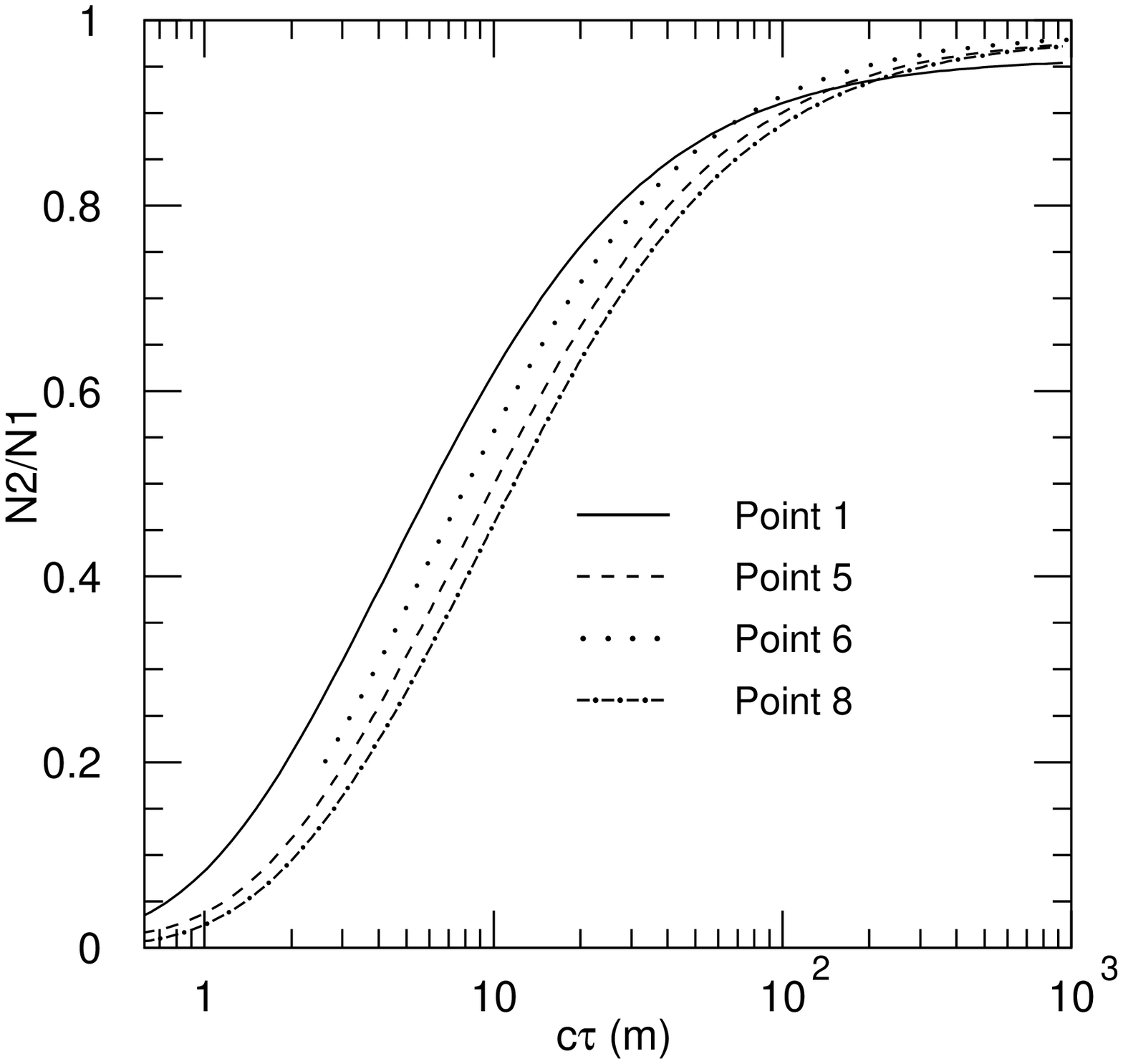}
}
\caption{\sl The ratio $R=N_2/N_1$ defined in the text as a function
of the slepton lifetime $c\tau$. Only the curves corresponding to
the model points 1, 5, 6, 8 are shown.}
\label{fig:ratio}
\end{figure}

The ratio:
$$
R=\frac{N_2}{N_1}
$$
is a function of the slepton lifetime. 
Its dependence on the
NLSP lifetime $c\tau$ in metres is shown in Fig.~\ref{fig:ratio}
for four of the eight model points. The curves for the model
points not shown are either very similar to one of the curves we
show or lie between the external curves corresponding to
models 1 and 8, thus providing no additional information.
Note that the curve for model 6 starts from $c\tau = 2.5$~m and not
from $c\tau = 50$~cm, as for the other models. This is due to the
large value of $M_{\rm mess}$ (cfr. Tab.~\ref{tab:tbg}), determining
a minimum NLSP lifetime allowed by theory which is macroscopic in
this case [cfr. Eqs.~(\ref{eq:NLSPtau}) and (\ref{eq:sqrtFmin})].

The probability $P$ for a particle of mass $m$ and momentum $p$
and proper lifetime $\tau$ to travel a distance $L$ before decaying is
given by the expression:
$$
P(L)=e^{-mL/p{\rm c}\tau}
$$
The value of $N_2$ is therefore a function of the momentum distribution
of the slepton, which is  determined by the details of the SUSY spectrum.
One needs therefore to simulate the full SUSY cascade decays
in order to construct the c$\tau-R$ relationship.

\begin{figure}[p]
\centerline{
\epsfxsize = 0.95\textwidth
\epsffile{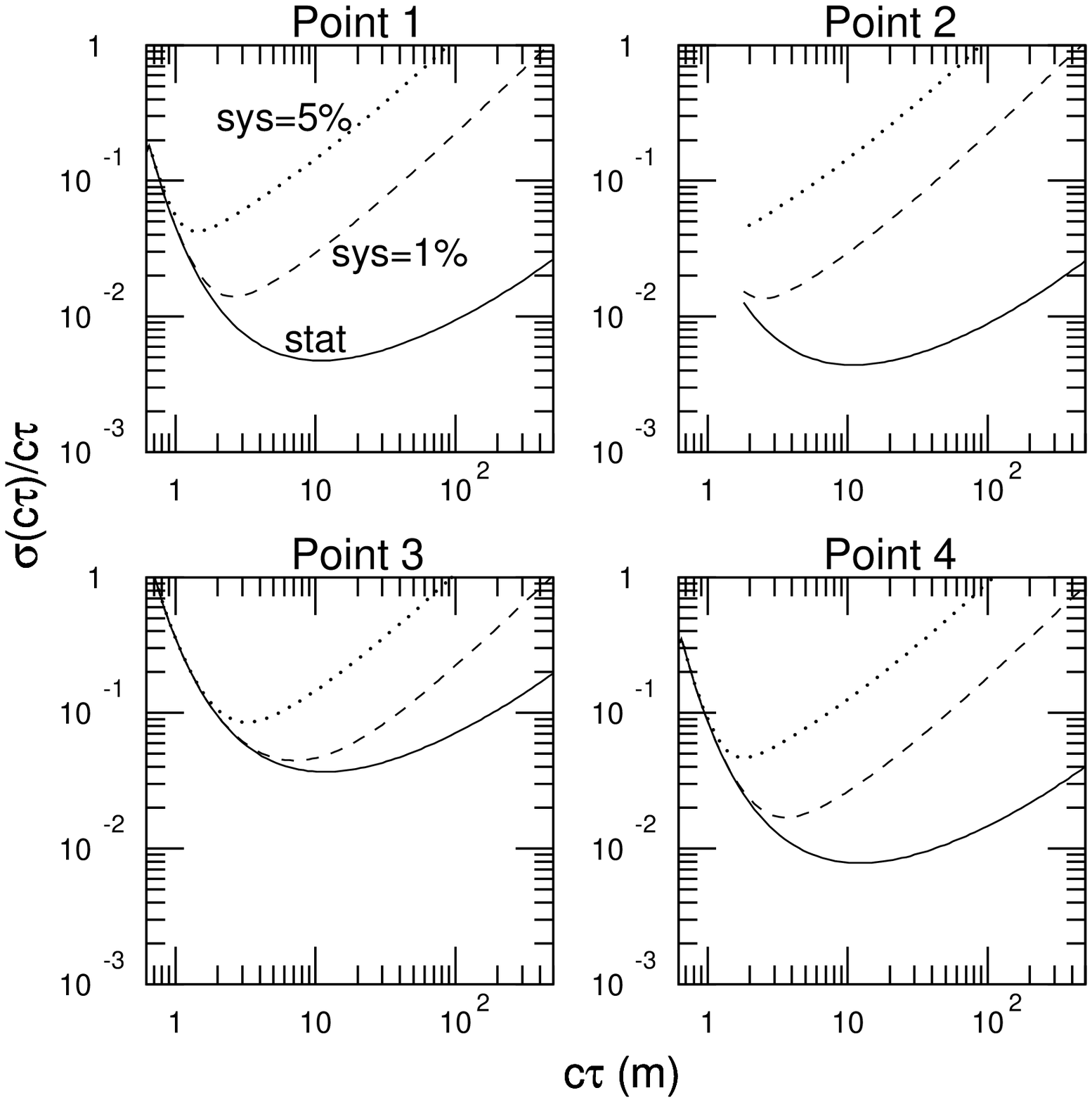}
}
\caption{\sl Fractional error on the measurement of the
slepton lifetime $c\tau$, for model points 1 to 4.
We assume an integrated luminosity of 30~fb$^{-1}$.
The curves are shown for three different assumptions on the
fractional systematic error on the $R$ measurement:
statistical error only (full line), 1\% systematic error (dashed line),
5\% systematic error (dotted line).
}
\label{fig:sigctau1}
\end{figure}

\begin{figure}[p]
\centerline{
\epsfxsize = 0.95\textwidth
\epsffile{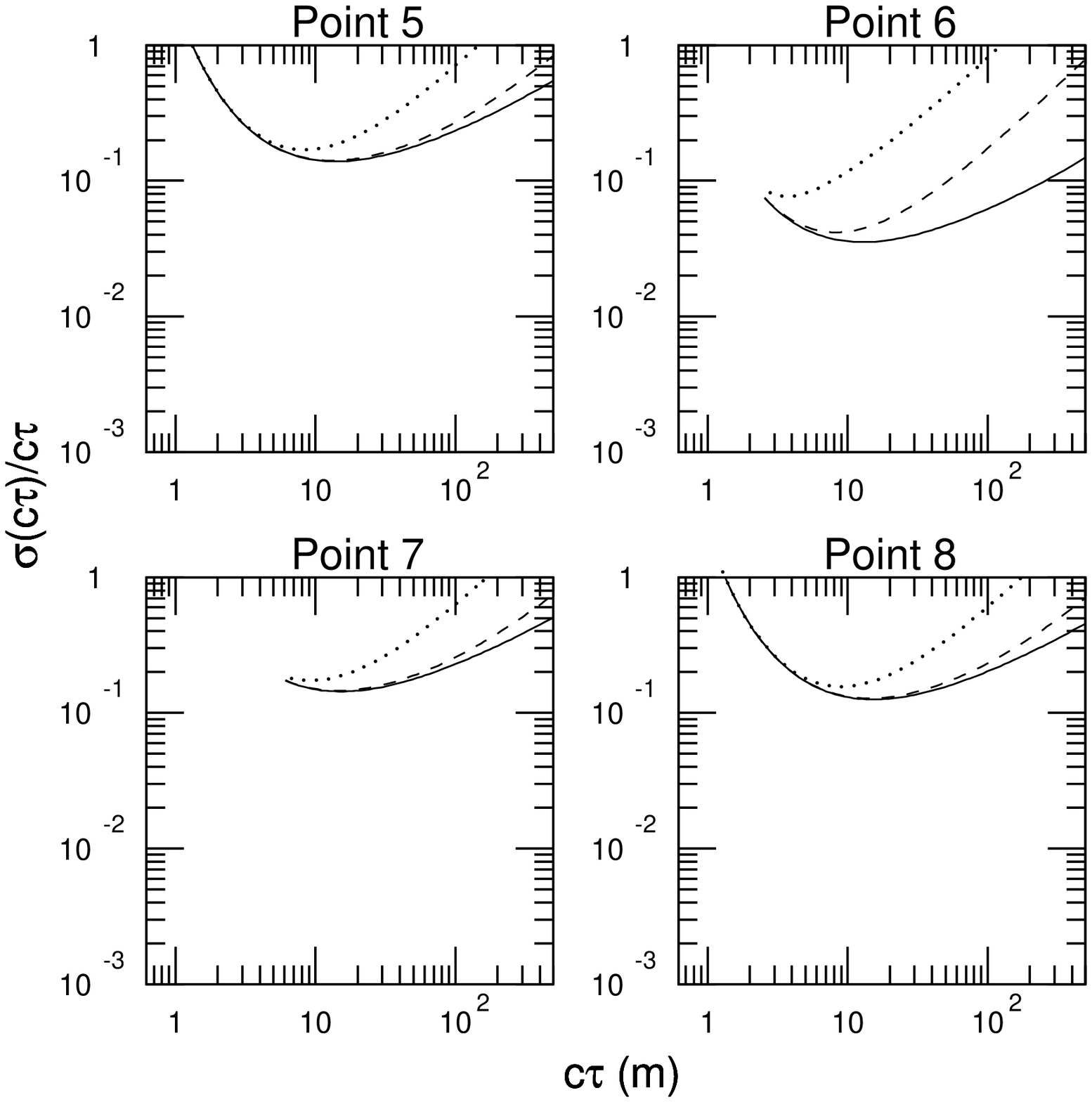}
}
\caption{\sl The same as in Fig.~\ref{fig:sigctau1}, but for
model points 5 to 8.
}
\label{fig:sigctau2}
\end{figure}

The statistical error in the $R$ measurement, can be evaluated as
$$
\sigma(R)=\sqrt{\frac{R\;(1-R)}{N_1}}
$$
Relevant for the precision with which the SUSY breaking 
scale can be measured is the error on the measured c$\tau$, which can be 
extracted from the curves shown in Fig.~\ref{fig:ratio}. 
This error can be evaluated as:
$$
\sigma({\rm c}\tau)=\sigma(R)/
\left[\frac{\partial R({\rm c}\tau)}{\partial{\rm c}\tau}\right]
$$
The measurement precision calculated according to this formula is shown 
in Figs.~\ref{fig:sigctau1} and \ref{fig:sigctau2} for the eight example
points, always for an integrated luminosity of 30~fb$^{-1}$. 
\def\topfraction{1.}
\def\textfraction{0.}

The full line in the plots is the error on c$\tau$ if only the statistical
error on $R$ is considered. The available statistics is a function 
of the mass scale of the strongly interacting sparticles. For 
mass scales between 500 and 1200~GeV, a statistical error smaller 
than 10\% can be achieved for c$\tau$ values ranging between 1~m and
several hundreds of metres. For a mass scale of  2000~GeV the statistical 
error is typically 10-20\%.
In the ideal case the details of the SUSY model are known and the 
$R-{\rm c}\tau$  relationship can be built from Monte Carlo, including
the effect of the detector acceptance. 
The subtraction of the background muons from the SUSY events is the 
dominant contribution to the systematic error on the $N_2$ measurement 
in this ideal case, and will be  treated in detail in the next section.\\ 
An additional uncertainty comes from the evaluation 
of the losses in $N_2$ because of sleptons produced outside
of the $\eta$ acceptance, or absorbed in the calorimeters, or  which escape 
the calorimeter with a transverse momentum below the cuts. This contribution 
is however expected to be much more important for the realistic case in 
which an imperfect knowledge of the SUSY model is assumed, and will 
be studied in that framework in a later section.\\
Since the uncertainty on R is a consequence of the uncertainty
on the evaluation of $N_2$, at this level we parametrise
the systematic error on R as a term proportional to R which is quadratically added
to the statistical error. We choose two values, \mbox{$1\%\;R$} and 
{$5\%\;R$}, and we propagate the error to the c$\tau$ measurement. 
The results are given as the dashed and the dotted lines in the plots 
in Figs.~\ref{fig:sigctau1} and \ref{fig:sigctau2}. 
\def\topfraction{1.}
\def\textfraction{0.}
For  squark mass scales up to 1200~GeV,  assuming a $1\%$ systematic error 
on the measured ratio, a precision better than 10\% on the c$\tau$ measurement
can be obtained for lifetime values between 0.5-1~m and 50-80~m. 
If the systematic uncertainty grows to 5\%, a 10\% precision
can only be achieved in the range 1-10~m. If the mass scale goes up to 2~TeV,
already at the level of pure statistical error a 10\% precision
is not achievable. One can however achieve a 20\% precision 
over c$\tau$ ranges between 5 and 100~m, if a 1\%  systematic 
error is assumed.

\subsection{Muon Background from SUSY Events}
\label{subsec:mubkgd}

\noindent
The definition of $N_2$, as described in the previous section, 
relies on a very loose identification of the second slepton 
candidate in order to minimise acceptance corrections, which are
the dominant source of systematic uncertainties on the $N_2/N_1$ ratio.

\begin{table}
\begin{center}
\begin{tabular}{|l||r|r|r|r|}
\hline
   &  \multicolumn{2}{c|}{No cut} & \multicolumn{2}{c|} 
{Cut Eq.~\ref{eq:2lep}}  \\
ID &  1 $\mu$ & $> 1\mu$ &    1 $\mu$ & $> 1\mu$  \\ \hline\hline
 1 & 0.37 & 0.14 & 0.027 & 0.00094 \\
 2 & 0.40 & 0.25 & 0.041 & 0.0011  \\
 3 & 0.23 & 0.42 & 0.22  & 0.043   \\
 4 & 0.39 & 0.39 & 0.16  & 0.014   \\
 5 & 0.22 & 0.42 & 0.10  & 0.0068  \\
 6 & 0.40 & 0.37 & 0.031 & 0.0013  \\
 7 & 0.36 & 0.34 & 0.026 & 0.      \\
 8 & 0.38 & 0.31 & 0.088 & 0.0040  \\
\hline
\end{tabular}
\end{center}
\caption{\sl Fraction of events with one or more detected muons for the eight
example points.}
\label{tab:nmu}
\end{table}

In the models considered for this study the charged sleptons are light,
and are mostly produced in the cascade decays of charginos 
and neutralinos through decays of the type $\tilde\chi^0\to\tilde \ell \ell$
and $\tilde\chi^\pm\to\tilde \ell \nu$.
Therefore in a significant fraction of the SUSY events muons will be produced
together with the sleptons. 
As can be seen from the second and third column of Tab.~\ref{tab:nmu},
at least one muon is produced in 60-80\% of the events for all considered 
models, and more than one muon in 30-40\% of the cases for most models.

\begin{figure}[p]
\centerline{
\epsfxsize = 0.9\textwidth
\epsffile{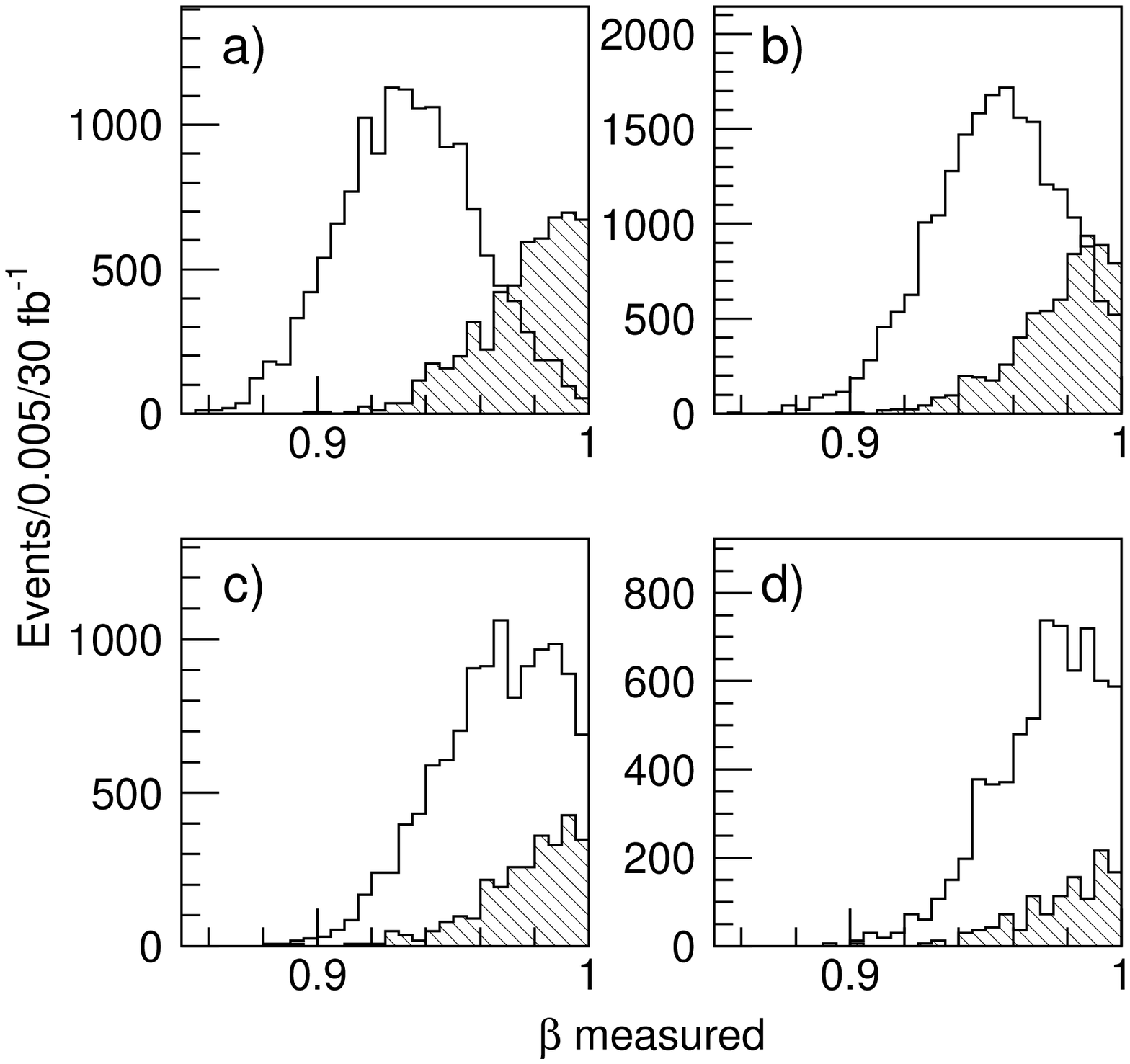}
}
\caption{\sl 
$\beta$ distributions for sleptons (full line histogram) and background
muons (hatched histogram) for model point 4. The distributions
are given for four momentum bins: a) $250<p<300$~GeV, 
$300<p<400$~GeV, $400<p<500$~GeV, $500<p<600$~GeV. The histogram limits 
exclude the peak at $\beta=1$ for the background.}
\label{fig:bdis}
\end{figure}

Most of these muons are soft, in particular in the case of 
$\tilde\tau_1$ NLSP, and are rejected by the requirement in the 
($P_{\rm meas}, \beta_{\rm meas}$) plane
which roughly corresponds to requiring the momentum of the
candidate to be above twice the slepton mass.  The number of muons
per events after this cut are given in columns four and five 
of Tab.~\ref{tab:nmu}. The background after cuts  is a function
of the mass difference between the squarks and the NLSP, 
which determines the lepton momentum spectrum, and of the NLSP 
mass which determines the minimum required momentum for a candidate.
The fraction ranges between a few percent and 20\% of the events. 

The background muons can be statistically subtracted using the 
observed $\beta$ distribution of the slepton candidates. 
For a fixed momentum $P$ of the candidate
the $\beta$ of the sleptons is peaked at the value 
$\beta=P/\sqrt{P^2+m^2}$ where $m$ is the slepton mass, whereas 
the distribution for the muons is peaked at $\beta=1$ and 
is essentially independent of the muon momentum.
As an illustration we show in Fig.~\ref{fig:bdis}
the $\beta$ distribution of sleptons (full line histogram)  and
of background muons (hatched histogram) for model point 4,
which has a high background contribution. The histogram 
limits are set to exclude the peak at $\beta=1$ to enhance the readability.
The distributions are given for four momentum bins, from 250 to 600~GeV,
and the clear difference in shape between signal and background
can be observed. 
The shape of the measured $\beta$ distribution for a given
momentum only depends on the mass of the particle, and 
will be known for both signal  and background  
both from detailed simulation of the detector response 
and from the data themselves. It will therefore be possible 
to measure bin by bin the relative contribution of signal and background
with a likelihood fit to the observed $\beta$ distributions. 
The signal/background separation obtained with this technique will  
be further validated by comparing the momentum spectrum 
of the background from isolated muons with the 
corresponding spectrum for electrons in SUSY events.

Given the good control of the detector response expected in ATLAS
and the multiple experimental handles available, 
the dominant contribution to the uncertainty on 
the measurement of $R$ from this source will be the statistical error 
on the background evaluation. This error will have a value
of approximately $\sqrt{k/N_1}$, where $k$ is the number in 
the fourth column of Table~\ref{tab:nmu}.
The contribution of this factor to the total error 
is only significant for the model points 3, 4, 5 and 8.
The main effect is a 30-40$\%$ degradation of the 
statistical error for c$\tau$ below 10~m for Models~3 and 8.
In all other cases, the effect on the c$\tau$ uncertainty from this source
is smaller than the curves labelled ``1\%'' in Figs.~\ref{fig:sigctau1} and
\ref{fig:sigctau2}.

\subsection{Model Independent Lifetime Measurement}
\label{subsec:modelind}

\noindent
Once the muon background has been subtracted, 
if the underlying SUSY model is known, $c\tau$ can simply
be measured from $R$ and the curves shown in Fig.~\ref{fig:ratio}.
Most of the SUSY mass spectrum will be measured 
from explicit reconstruction of exclusive decay chains, as shown in 
\cite{ihfp} and\cite{TDR}. It is difficult however at this stage to evaluate
the uncertainty in the construction of the $R$-c$\tau$ calibration curve
from an imperfect knowledge of the SUSY model.
As an alternative approach, the measurement can be performed 
by deconvoluting the effect of the relativistic lifetime 
dilatation from the measured momentum distribution of the slepton candidates.
There are two cases to consider for each event, depending on whether both slepton
candidates pass the cuts used to define $N_1$ or only one does. 
If only one of the two slepton candidates passes the $N_1$ cuts, 
the basic equation is:
\begin{equation}
N_1={\cal C}\sum_{i=1}^{N_2}W_i,
\label{eq:mid}
\end{equation}
with
$$
W_i=e^{\frac{Lm}{c\tau P_{2,i}}},
$$
where $L$ is the distance of the outermost muon station, $m$ is the slepton 
mass, $P$ its momentum, and the subscript 2 refers to the $i$-th slepton 
candidate which did not pass the cuts used to define $N_1$. 
The acceptance correction ${\cal C}$ is defined as the reciprocal of the 
detector acceptance and the experimental efficiency in detecting
the second slepton candidate. Its value is strictly greater than 1.\\
The fraction of events in which both legs pass the criteria to define 
$N_1$ varies between $\sim$15\% and 40\%, increasing with the NLSP mass. 
In this case the expression for the event weight $W_i$ must be symmetrised to:
$$
W_i=\frac{e^{-\frac{Lm}{c\tau P_{1,i}}}+e^{-\frac{Lm}{c\tau P_{2,i}}}}
{e^{-\frac{Lm}{c\tau}\left(\frac{1}{P_{1,i}}+\frac{1}{P_{2,i}}\right)}} - 1
$$

\begin{figure}[t]
\centerline{
\epsfxsize = 0.7\textwidth
\epsffile{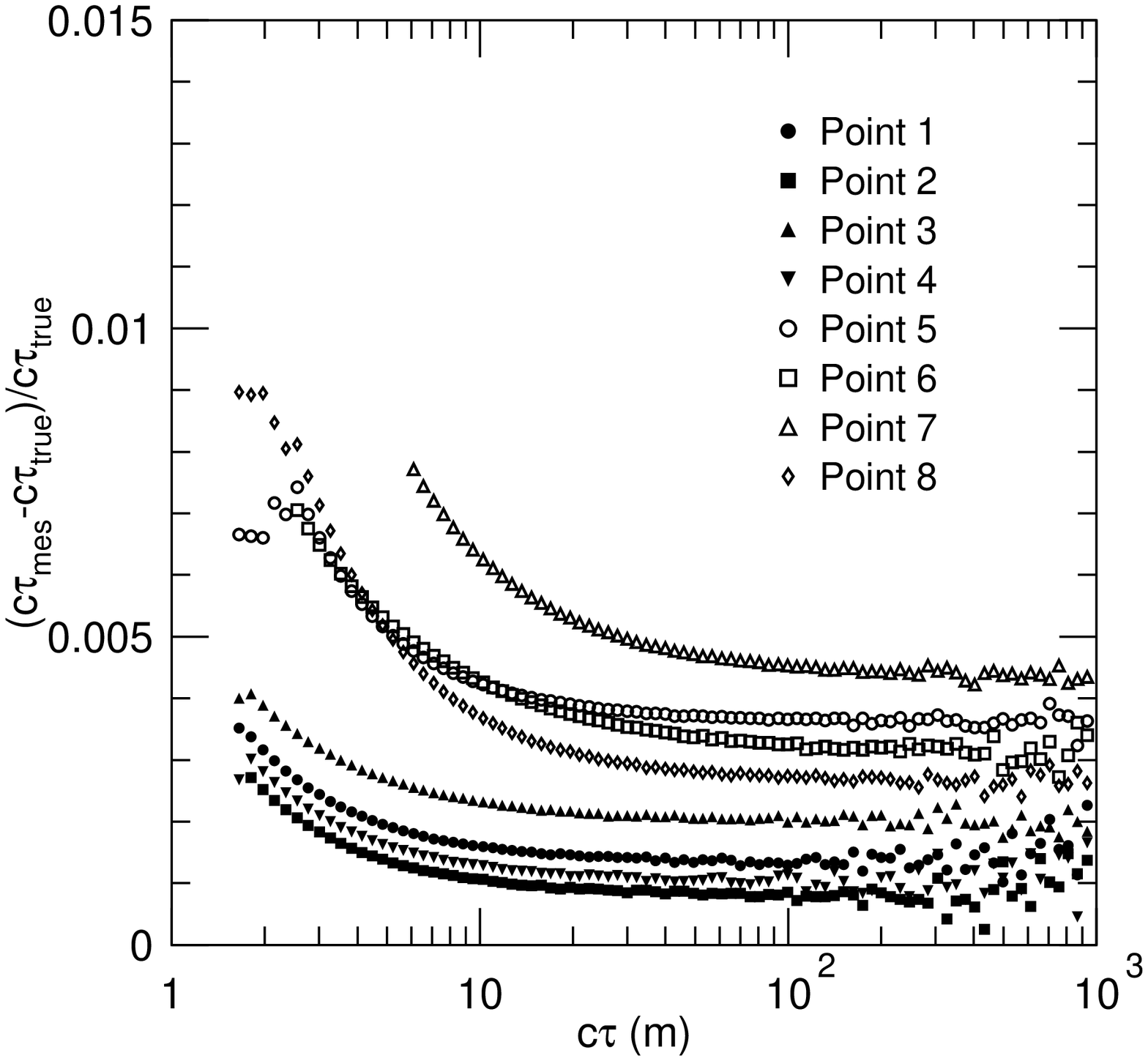}
}
\caption{\sl Fractional deviation of the c$\tau$ value measured
using the momentum distribution of the candidates from the true
value as a function of c$\tau$. The different curves shown are
for the 8 example points considered. The statistical error and 
the systematic uncertainty on $N_2$ are not 
included in the calculation.
}
\label{fig:method}
\end{figure}

By solving Eq.~\ref{eq:mid}, the value of c$\tau$ can be 
measured with no reference to the underlying model.
In order to evaluate the accuracy of the method, we measure 
c$\tau$ from the momentum distribution of the candidate sleptons
for different input c$\tau$ values, assuming no statistical error and 
no systematic uncertainty from muon background subtraction,
or from the evaluation of the acceptance correction ${\cal C}$.
The fractional error on the measured c$\tau$ is shown in Fig.~\ref{fig:method} 
as a function of c$\tau$ for all the 8 model points,
under the above assumptions.
For all the points and for the range 5--1000~m
the deviation of the calculated c$\tau$ from the real value is less than 1\%.
The source of this small systematic deviation is the momentum smearing of 
the sleptons, which causes a few sleptons to be lost because they fall 
below the analysis cuts, and the fact that only an average correction is 
applied to compensate for ionisation energy loss in the calorimeters.

The experimental sources of uncertainty 
in Equation~\ref{eq:mid}, except the $p$-scale,
can be parametrised as a deviation
of ${\cal C}$ from the true value. The propagation of the uncertainty 
on ${\cal C}$ to the c$\tau$ measurement was explicitly studied 
by calculating c$\tau$ with our  simulated samples for all the accessible 
c$\tau$ range using a value of ${\cal C}$ shifted by 1$\%$ 
with respect to the true value. The resulting displacement
in the calculated c$\tau$ value is accurately described by the curves
labelled "1\%" in Figures~\ref{fig:sigctau1} and \ref{fig:sigctau2}.

At this point, to complete the evaluation of the systematic uncertainty 
on c$\tau$, we need to discuss how the acceptance correction ${\cal C}$ 
can be estimated in the ATLAS detector, and the expected uncertainty
on its value.

\subsection{Systematic Uncertainties on acceptance}
\label{subsec:systematics}

\noindent
The main experimental effects causing the loss of a slepton 
produced in a GMSB event are:
\begin{itemize}
\item
the low energy sleptons which due to the ionisation energy loss 
in the calorimeters fall below
the energy and transverse momentum requirements of the analysis; 
\item
the $|\eta|$ acceptance of the detector.
\end{itemize}
An additional $\sim$5\% loss, coming from 
the reconstruction efficiency will be measured with 
high precision exploiting the redundancy of the various
ATLAS subdetectors, and will not be further considered.

The loss of low momentum sleptons can be estimated by 
studying the spectrum of the sleptons which 
range out in the hadronic calorimeter. These particles
should present the characteristic signature of a stiff isolated
highly ionising track in the inner detector depositing 
a small amount of energy in the electromagnetic calorimeter
and all of its kinetic energy in a single tower of  the hadronic 
calorimeter.
A detailed study with full simulation of the ATLAS detector 
is needed to assess how well the acceptance loss can be evaluated
with this technique. 

For the study of the $|\eta|$ acceptance no such clear handle exists,
but it should be possible to extract some indications from the
observed $\eta$ distribution, and by studying the tracks  
up to a pseudorapidity $|\eta|<$2.7 which is the 
limit of the acceptance of the precision muon chambers.

The acceptance correction can also be evaluated 
through a Monte Carlo simulation of all the SUSY processes.
The SUSY events are dominated by the production of
squarks and gluinos, and the $\eta$ distribution of the sparticles
produced in the hard scattering is a function
of their mass. In the considered models
the significant mass difference between squarks and gluinos
and their decay products produces rather collimated decays,
and the $\eta$ distribution of the NLSP's is mostly 
determined by the mass scale of squarks and gluinos.
On the other side, the lower end of the NLSP momentum 
spectrum is dominated by the direct production  of 
sleptons charginos and neutralinos.

\begin{figure}[t]
\centerline{
\epsfxsize = 0.7\textwidth
\epsffile{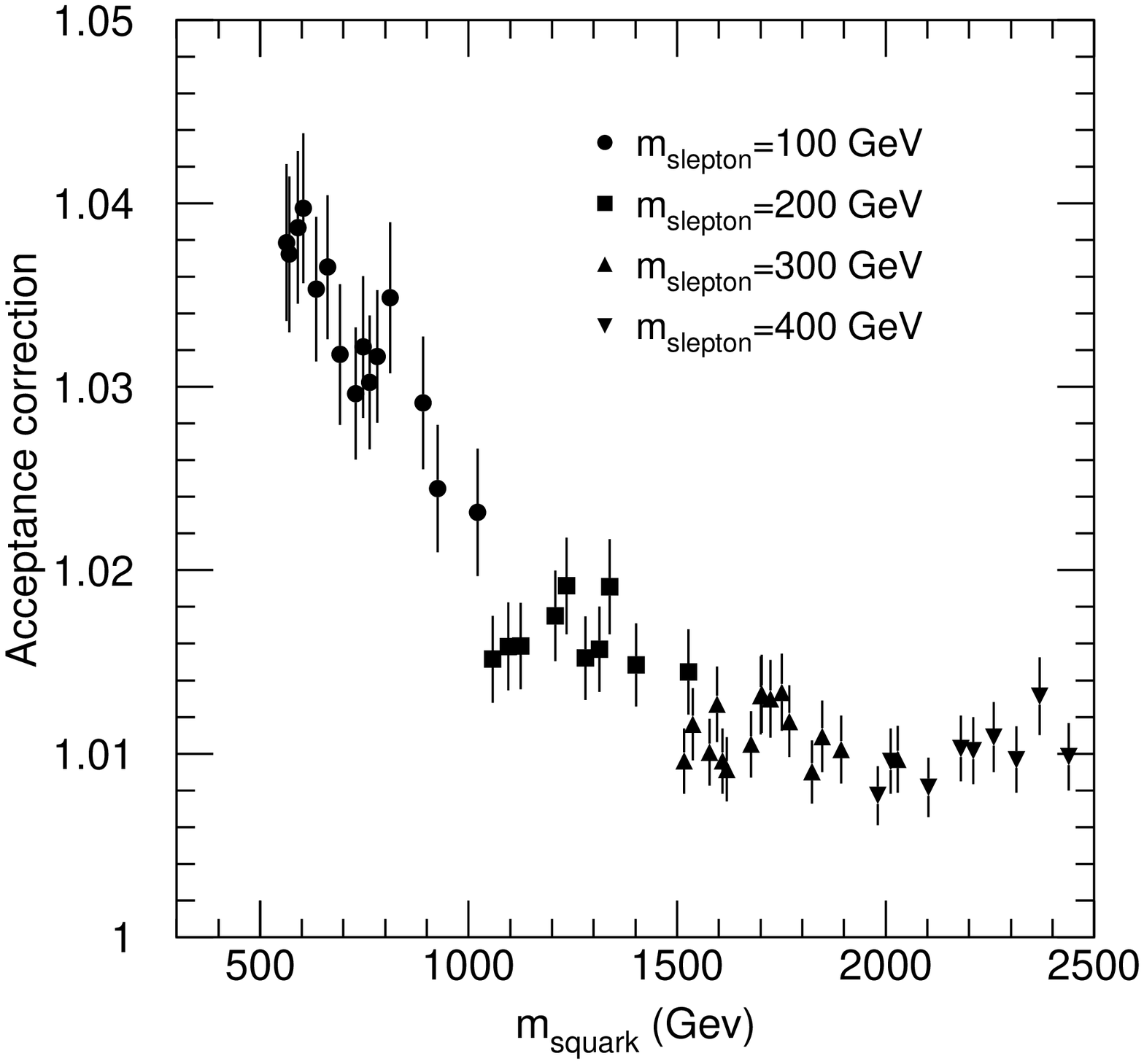}
}
\caption{\sl Acceptance correction factor calculated for 50 different
GMSB models. The effects of the loss
of low momentum sleptons and of the $\eta$ acceptance are included.
The acceptance is given as a function of the squark mass.
}
\label{fig:acc}
\end{figure}

In order to evaluate the spread in the value of the acceptance  
for different model assumptions, we have analysed 50 models
with NLSP masses of 100, 200, 300, or 400~GeV, and  a spread as large as 
possible in squark mass scale. We show in Fig.~\ref{fig:acc}
the acceptance correction ${\cal C}$ for the 
50 models as a function of the squark mass scale, calculated as the average 
of the masses of all the six squark flavours, both left and right handed. 
The correction varies between 4\% and 1\% with increasing squark mass
levelling at 1\% for squark masses higher than 1500~GeV. The spread 
in the correction factor for a fixed squark mass is below 1\%. 
For a general SUSY model, the squark mass will be known, from 
the inclusive study  of the $E_T$ distribution in SUSY events
to 5-10\% \cite{TDR}. The situation is even better 
in the GMSB scenario addressed in this study, for which it will 
be possible to perform the full reconstruction
of the decay chains of squarks \cite{ihfp}, yielding an error on squark masses
at the percent level.
From these considerations, even if we conservatively assume only 
the constraints from inclusive studies, it should be possible 
to keep the uncertainty on the acceptance correction ${\cal C}$ 
well below the 1\% mark. 

\section{Determining the SUSY Breaking Scale $\sqrt{F}$}
\label{sec:sqrtF}

\noindent
Using the measured values of $c\tau$ and the NLSP mass,
the SUSY breaking scale $\sqrt{F}$ can be calculated from
Eq.~(\ref{eq:NLSPtau}), where ${\cal B} = 1$ for the case where
the NLSP is a slepton.
From simple error propagation, the fractional uncertainty on the
$\sqrt{F}$ measurement can be obtained from the experimental uncertainties
on $c\tau$ on the slepton mass.

\begin{figure}[p]
\centerline{
\epsfxsize = \textwidth
\epsffile{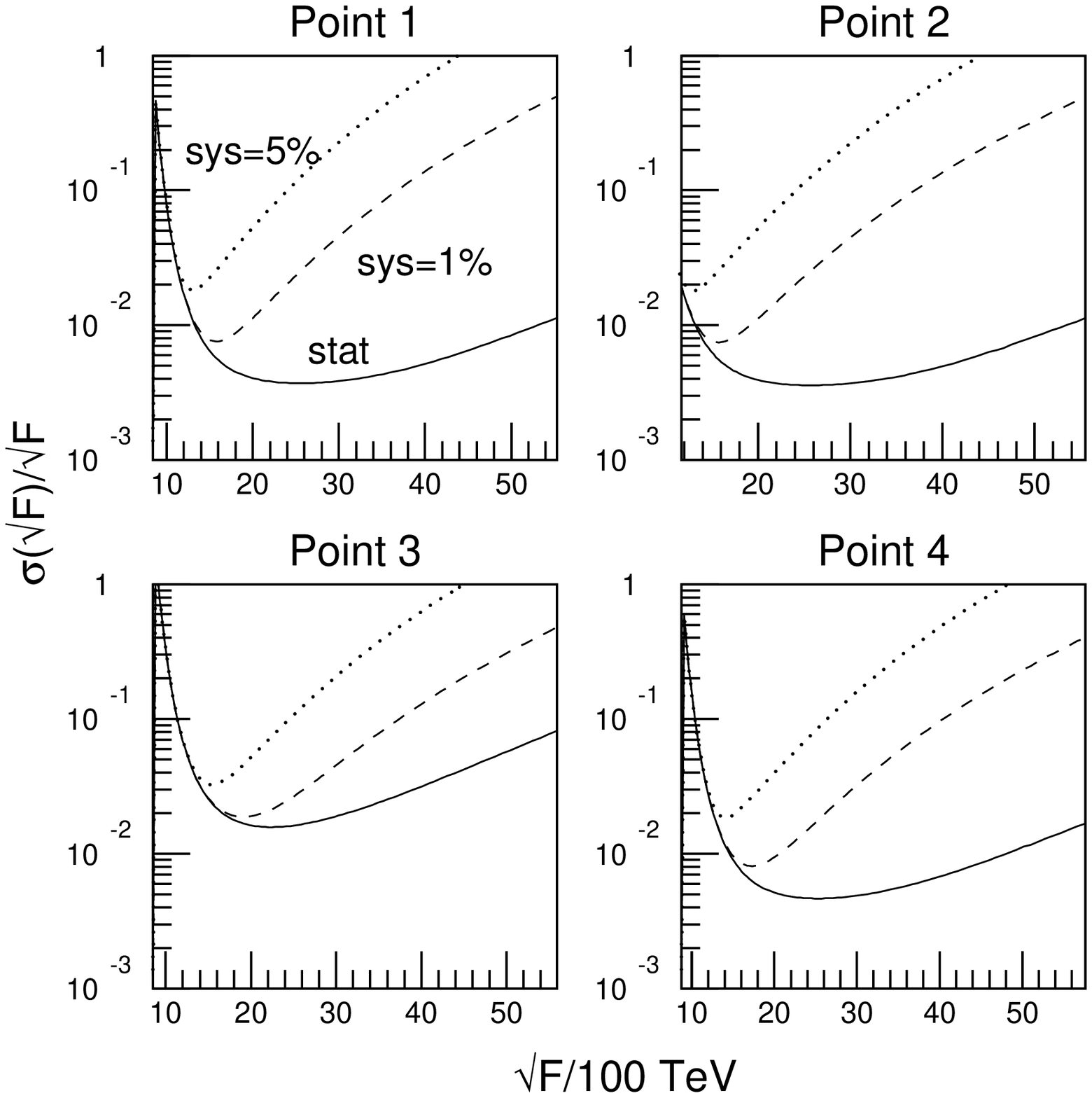}
}
\caption{\sl Fractional error on the measurement of the
SUSY breaking scale $\sqrt{F}$ for model points 1 to 4.
We assume an integrated luminosity of 30~fb$^{-1}$.
The curves are shown for the three different assumptions on the fractional
systematic error used in Figs.~\ref{fig:sigctau1} and \ref{fig:sigctau2}.
}
\label{fig:sigsqrtf1}
\end{figure}

\begin{figure}[p]
\centerline{
\epsfxsize = \textwidth
\epsffile{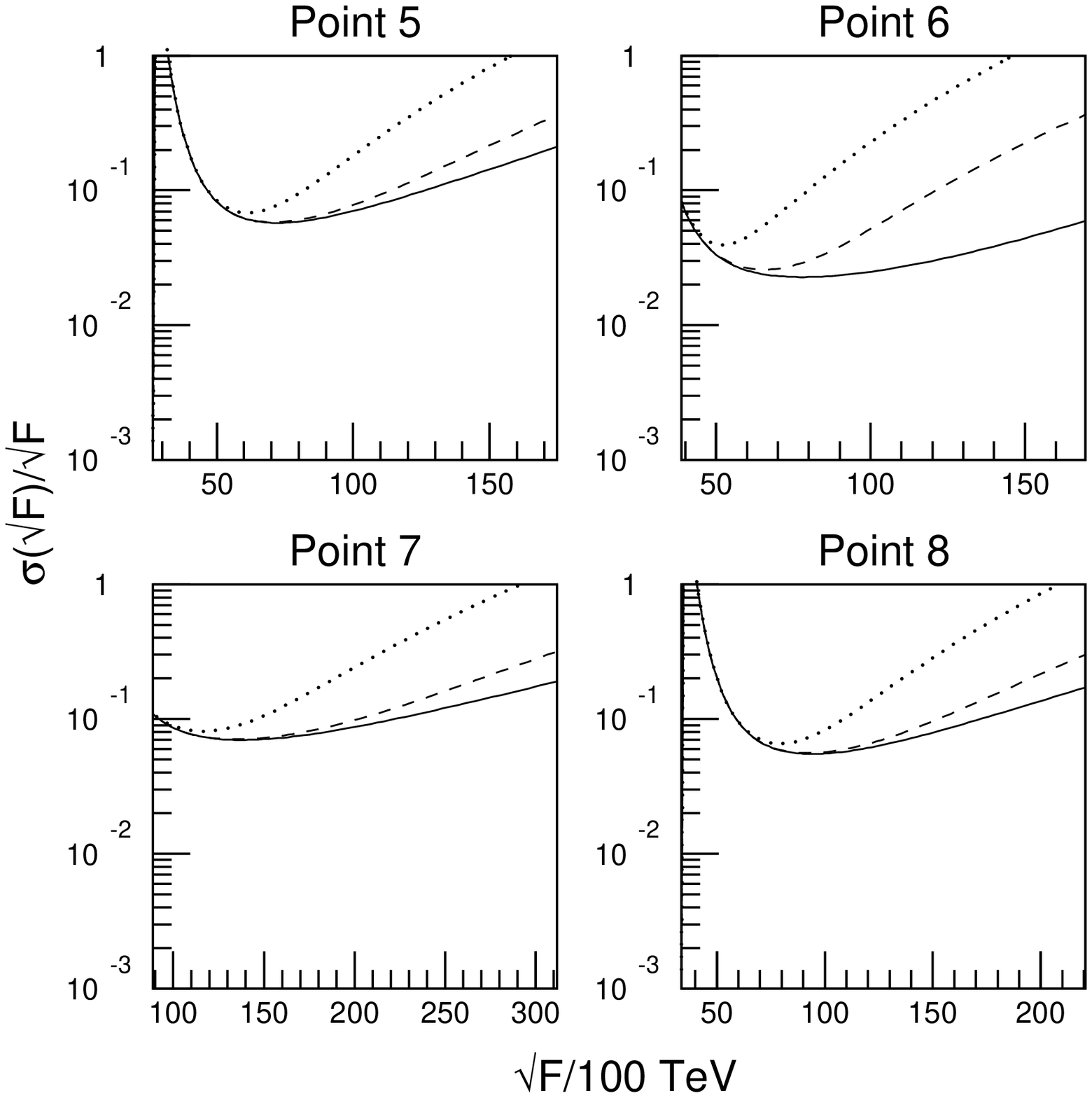}
}
\caption{\sl The same as in Fig.~\ref{fig:sigsqrtf1}, but for 
model points 5 to 8.
}
\label{fig:sigsqrtf2}
\end{figure}

In Figs.~\ref{fig:sigsqrtf1} and \ref{fig:sigsqrtf2}, we show the fractional
error on the $\sqrt{F}$ measurement as a function of $\sqrt{F}$ for
our three different assumptions on the $c\tau$ error.
The uncertainty is dominated by $c\tau$ for the higher part of the
$\sqrt{F}$ range and grows quickly when approaching
the lower limit on $\sqrt{F}$. This is because very few sleptons
survive and the statistical error on both $m_{\tilde \ell}$ and
$c\tau$ gets very large.
If we assume a 1\% systematic error on the ratio $R$ from which $c\tau$
is measured (dashed lines in Figs.~\ref{fig:sigsqrtf1} and
\ref{fig:sigsqrtf2}), the error on $\sqrt{F}$ is better than 10\%
for $1000 \ltap \sqrt{F} \ltap 4000$~TeV for model points 1--4 with
higher statistics. For points 5--8, in general one can explore
a range of higher $\sqrt{F}$ values with a small relative error,
essentially due to the large NLSP mass in these models.
Note also that the theoretical lower limit (\ref{eq:sqrtFmin}) on $\sqrt{F}$
is equal to about 1200, 1500, 3900, 8900 TeV respectively in model
points 2, 5, 6, 7, while it stays well below 1000 TeV for the other models.

\section{Conclusions}
\label{sec:conc}

\noindent
We have discussed a simple method to measure at the LHC with the ATLAS
detector the fundamental SUSY breaking scale $\sqrt{F}$ in the GMSB scenarios
where a slepton is the NLSP and decays to the gravitino with a lifetime
in the range 0.5~m $\ltap c\tau_{\rm NLSP} \ltap 1$~km. This method requires
the measurement of the time of flight of long lived sleptons
and is based on counting events with one or two identified NLSP's.
The achievable measurement precision critically depends on the
uncertainties in evaluating the experimental inefficiencies 
in the NLSP detection.
We have performed a particle level simulations for eight representative
GMSB models, some of them being particularly hard due to low statistics.
The experimental study is based on a parametrisation of 
the ATLAS muon detector response to sleptons, based on a detailed full 
simulation study. The careful consideration of the possible 
sources of uncertainty allows us to conclude that the systematic
uncertainty affecting the measurement will be at the percent level.
In this framework, a level of precision of a few 10's \% on the SUSY breaking
scale measurement can be achieved in significant parts of the
$1000 \ltap \sqrt{F} \ltap 30000$~TeV range, for all models considered.
The range of the measurement could be extended through the direct detection
of NLSP decays either inside the inner detector cavity or inside 
the muon spectrometer. A detailed detector simulation for these signatures 
would be needed in order to assess the possible gain in sensitivity.

We stress that the results of the present analysis cover larger classes of 
theoretical frameworks. In particular, any model implying the presence of 
long-lived particles decaying into leptonic final states through the 
production of primary heavy particles with mass of order 1 TeV        
(parameter that guarantees a crucial suppression of the SM background)
can be analysed according to similar strategies. 

\vspace{1.0cm}

\subsection*{Acknowledgements}
We would like to acknowledge useful discussions with many members of the ATLAS
Collaboration about the details of the ATLAS experimental setup.
Special thanks go to Frank Paige for discussions on the signatures of
the slow sleptons and for useful comments on the paper, and to Fabiola 
Gianotti for carefully reading the manuscript and for punctual
remarks on the experimental assumptions underlying our analysis.
S.~A. and G.~P. thank the organisers of the Workshop ``Physics at TeV
Colliders'', for the hospitality and the pleasant and productive atmosphere
in Les Houches where this work was started.


\vspace{0.7cm}

\begin{thebibliography}{10}

\bibitem{StevePrimer}
  For a recent pedagogical review of supersymmetry and supersymmetry 
  breaking, see S.~P.~Martin, ``A Supersymmetry Primer'', in ``Perspectives
  on Supersymmetry'', G.~L.~Kane ed., World Scientific 1998, hep-ph/9709356 
  and references therein. 

\bibitem{oldGMSB}
    M.~Dine, W.~Fischler, M.~Srednicki, \NPB{189}{1981}{575};
    S.~Dimopoulos, S.~Raby, \NPB{192}{1981}{353};
    M.~Dine, W.~Fischler, \PLBold{110}{1982}{227};
    M.~Dine,  M.~Srednicki, \NPB{202}{1982}{238};
    M.~Dine, W.~Fischler, \NPB{204}{1982}{346};
    L.~Alvarez-Gaum\'e, M.~Claudson, M.~B.~Wise, \NPB{207}{1982}{96};
    C.~R.~Nappi, B.~A.~Ovrut, \PLBold{113}{1982}{175};
    S.~Dimopoulos, S.~Raby, \NPB{219}{1983}{479}.

\bibitem{newGMSB}
    M.~Dine, A.~E.~Nelson, \PRD{48}{1993}{1277};
    M.~Dine, A.~E.~Nelson, Y.~Shirman, \PRD{51}{1995}{1362};
    M.~Dine, A.~E.~Nelson, Y.~Nir, Y.~Shirman, \PRD{53}{1996}{2658}.

\bibitem{GR-GMSB}
  For a review, see G.~F.~Giudice, R.~Rattazzi, \PREP{322}{1999}{419}.

\bibitem{GMSBmodels1}
    S.~Dimopoulos, S.~Thomas, J.~D.~Wells, \PRD{54}{1996}{3283};
    \NPB{488}{1997}{39}.

\bibitem{GMSBmodels2}
    J.~A.~Bagger, K.~Matchev, D.~M.~Pierce, R.~Zhang, 
    \PRD{55}{1997}{3188}. 

\bibitem{AKM-LEP2} 
    S.~Ambrosanio, G.~D.~Kribs, S.~P.~Martin, \PRD{56}{1997}{1761}. 

\bibitem{AB-LC}
    S.~Ambrosanio, G.~A.~Blair, \EPJC{12}{2000}{287--321}.

\bibitem{Fayet}
    P.~Fayet, \PLBold{70}{1977}{461}; \PLBold{86}{1979}{272};
    \PLB{175}{1986}{471} and in ``Unification of the fundamental 
    particle interactions", eds.~S.~Ferrara, J.~Ellis,   
    P.~van Nieuwenhuizen (Plenum, New York, 1980) p.~587.

\bibitem{AKM2} 
  A relevant example is discussed in:
  S.~Ambrosanio, G.~D.~Kribs, S.~P.~Martin, \NPB{516}{1998}{55}.

\bibitem{TDR}
  The ATLAS Collaboration, ``ATLAS Detector and Physics Performance Technical
  Design Report'', ATLAS TDR 15, CERN/LHCC/99-15 (1999).

\bibitem{SUSYFIRE}
  An updated, generalised and {\tt Fortran}-linked version of the 
  program used in Ref.~\cite{AKM-LEP2}. It generates minimal and 
  non-minimal GMSB and SUGRA models.
  For inquiries about this software package, please send e-mail to  
 {\tt ambros@mail.cern.ch}. 

\bibitem{isajet}
  H.~Baer, F.~E.~Paige, S.~D.~Protopopescu, X. Tata, hep-ph/9305342; 
  hep-ph/9804321.

\bibitem{ATLFAST}
  E.~Richter-Was, D.~Froidevaux, L.~Poggioli, ``{\tt ATLFAST 2.0}: A Fast
  Simulation Package for ATLAS'', ATLAS Internal Note ATL-PHYS-98-131 (1998).

\bibitem{leandro}
  A.~Nisati, S.~Petrarca, G.~Salvini, Mod. Phys. Lett. {\bf A12} (1997) 2213.

\bibitem{drtata}
  M.~Drees, X.~Tata, \PLB{252}{1990}{695}.

\bibitem{femoroi}
  J.~L.~Feng, T.~Moroi, \PRD{58}{1998}{035001}.

\bibitem{marthom}
  S.~P.~Martin, J.~D.~Wells, \PRD{59}{1999}{035008}.

\bibitem{leandro1}
  A.~Nisati, ``Preliminary Timing Studies of the Barrel Muon Trigger 
  System'', ATLAS Internal Note ATL-DAQ-98-083 (1998).

\bibitem{gpar}
  G.~Polesello, A.~Rimoldi, ``Reconstruction of Quasi-stable Charged
  Sleptons in the ATLAS Muon Spectrometer'', 
  ATLAS Internal Note ATL-MUON-99-06.

\bibitem{ihfp}
  I.~Hinchliffe, F.~E.~Paige, \PRD{60}{1999}{095002}.

\bibitem{PYTHIA} 
  T.\,Sj\"ostrand, Comp. Phys. Comm. {\bf 82} (1994) 74. 

\end{thebibliography}
\end{document}